\documentclass{aa}
\usepackage{amsmath}
\usepackage{graphicx}
\usepackage{epsfig}
\usepackage{changebar}
\usepackage{natbib}
\usepackage{multirow}
\newcommand{\HII}{\mbox{HII}}

\newcommand{\kms}{km s$^{-1}$}
\newcommand{\cmthree}{cm$^{-3}$}
\newcommand{\cmtwo}{cm$^{-2}$}
\newcommand{\ks}{\emph{K$_s$}}
\newcommand{\Jm}{\emph{J}}
\newcommand{\Hm}{\emph{H}}
\newcommand{\x}{$\times$}

\begin{document}
\title {High-mass star formation within the \mbox{bright-rimmed cloud SFO 79}}
\author{J. S. Urquhart\inst{1}, M. A. Thompson\inst{1,2}, L. K. Morgan\inst{1} \& Glenn J. White\inst{1}}
\offprints{J. S. Urquhart: jsu@star.kent.ac.uk}
\institute{Centre for Astrophysics and Planetary Science, School of Physical Sciences, University of 
Kent, Canterbury, CT2 7NR, England
\and 
Centre for Astrophysics Research, Science and Technology Research Institute,
University of Hertfordshire, College Lane, Hatfield, AL10 9AB, England}
\abstract{We report Radio Recombination Line (RRL) and continuum observations toward the IRAS point source
16362--4845, embedded within the Bright-Rimmed Cloud (BRC) SFO 79, a small molecular cloud lying at the 
edge of the HII region RCW 108. High resolution observations of the H92$_\alpha$ hydrogen recombination 
line and of the continuum emission (3.6 and 6 cm) confirm the presence of a resolved Ultra Compact (UC) HII 
region embedded within the molecular cloud. The integrated radio fluxes suggest the source of the 
ionisation to be an O9 Zero Age Main Sequence (ZAMS) star. Millimetre observations of  $^{12}$CO, 
$^{13}$CO and C$^{18}$O (\emph{J}=1--0) molecular lines reveal the presence of a molecular condensation 
offset $\sim$30\arcsec~to the north of the IRAS position on the boundary of the UC HII region. Analysis of
2MASS data has led to the identification of a small IR cluster of Young Stellar Objects (YSOs) that are 
positionally coincident with the UC HII region, lying to the south east of the peak of the radio emission. 
Moreover, the UC HII region appears to be extended in the direction of the IR cluster, which suggests that 
the radio emission and the IR cluster are in some way related to each other. MSX 8.3 $\mu$m and 21.3 
$\mu$m images have been used to trace the large scale structure of the BRC, revealing the presence of a 
Photo Dominated Region (PDR) and three embedded thermal sources within the molecular cloud. The PDR has a 
plane parallel morphology which correlates extremely well with the morphology of the ionised gas traced by 
the optical emission. The three thermal sources (labelled A, B, C) all lie at a similar projected distance 
from the interface between the HII region and the molecular gas of the cloud. Thermal sources A and C are 
positionally coincident with the IRAS point sources 16362--4845 and 16362--4841 respectively, both of which
have IRAS colours consistent with the presence of UC HII regions. Given that UC HII regions are relatively 
short lived ($\sim$10$^5$ yrs) it is reasonable to suggest that these two UC HII regions are of a similar 
age. The alignment of the three thermal sources along a line parallel to the bright rim suggests that they 
could have been triggered by the propagation of a plane parallel shock through the cloud.
}
\authorrunning{J. S. Urquhart et al.}
\titlerunning{Massive star formation within SFO 79}
\maketitle
\section{Introduction}
Bright-Rimmed Clouds (BRCs) are small molecular clouds associated with evolved HII regions, and have long 
been considered possible sites for star formation triggered by the expansion of the surrounding HII 
region. The far-UV photons emitted from an OB star, located toward the centre of the HII regions, ionise 
the surface of the molecular cloud and drive photoionisation-induced shocks into the molecular material. 
These shocks compress the molecular gas, forming dense cores and possibly triggering their collapse 
as well as triggering the collapse of any pre-existing cores that may already be present within the 
molecular cloud in a process known as Radiative-Driven Implosion (RDI; 
\citealt{bertoldi1989, bertoldi_1990, lefloch1994}, 1995).

Sugitani, Fukui, \& Ogura (1991) identified 44 BRCs in the northern hemisphere which are associated with
IRAS point sources, suggesting the presence of embedded protostars. This was followed by a search 
of the southern sky for BRCs, using the ESO(R) Southern Hemisphere Atlas by \citet{sugitani1994}.
This search identified a further 45 BRCs associated with IRAS point sources. In total 89 BRCs have been
identified as potential sites for triggered star formation to occur, collectively known as the SFO 
catalogue. 

A recent radio continuum survey of 45 southern BRCs between 3.6 and 20 cm (Thompson, Urquhart \& White, 
2004a; hereafter Paper I) detected five radio sources that, because of their positional coincidence with 
embedded IRAS point sources, were considered as possible compact HII regions. Compact HII regions are 
signposts pointing to the earliest stages of massive star formation in which newly formed stars are still 
embedded within their natal molecular clouds and are ionising the surrounding gas. The BRC SFO 79 was 
identified in Paper I as a possible high-mass star forming site and is the subject of this study. The high 
infrared and radio luminosity detected toward the embedded IRAS point source 16362--4845 within this BRC 
suggest the presence of an embedded stellar cluster surrounded by a compact HII region driven by a late O 
type star. This embedded source has also been associated with an infrared cluster found in 2MASS survey 
data (Paper I; \citealt{dutra_2003}). In order to confirm the nature of star formation within this BRC 
we have made high resolution observations of the possible compact HII region detected in our earlier 
survey.

\begin{figure*}[!tbp]
\begin{center}
\epsfig{file=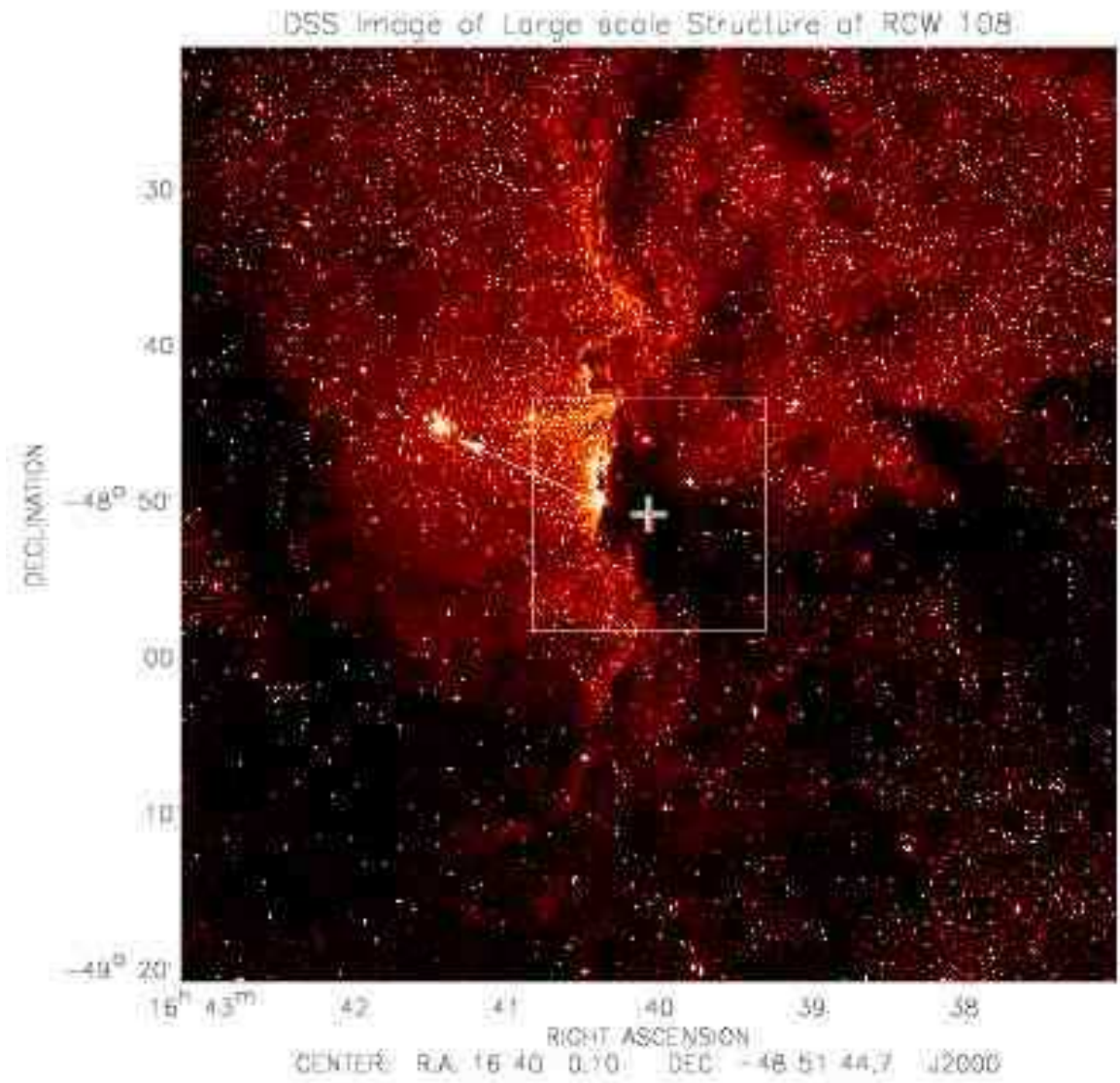,height=16cm,width=16cm, clip=}
\caption{ DSS (R band) image of the HII region RCW 108 covering an area of one square degree. The location
of SFO 79 is outlined by the white square in the centre of the image. The centre cross indicates the 
position of the embedded IRAS point source 16362-4845. Ionising stars lie to the east of the optical 
bright-rim in the direction of the white arrow.}
\label{fig:rcw108}
\end{center}
\end{figure*}

SFO 79 is a small molecular cloud on the edge of RCW 108, a bright HII region located at a heliocentric 
distance of $\sim$~1.35 kpc (\citealt{georgelin1996}). This cloud has been associated with the IRAS point 
source 16362--4845, which has the IRAS colours consistent with the presence of an embedded protostar or 
an UC HII region. RCW 108 is excited by two O stars, HD 150135 and HD 150136, located at a projected 
distance of 4.2 pc east from the edge of the bright rim of the cloud (\citealt{whiteoak1963, herbst1977}; 
see Figure~\ref{fig:rcw108}). The bright rim of the cloud marks the interface between the HII region and 
the dense dark material. These two stars are part of NGC 6193 forming the core of the Ara OB1 association 
\citep{whiteoak1963}. The coincidence of the embedded IRAS point source with the several reflection 
nebulae that form the R association Ara R1 (Herbst 1974, 1975) has brought the question of active star 
formation in these molecular clouds to the fore. Since the detection of the H109$_\alpha$ RRL 
\citep{wilson1970} toward the centre of the cloud SFO 79 has been the focus of some interest. 
\citet{wilson1970} reported the detection of a small unresolved region with an emission measure of 
4.1$\times10^{6}$ cm$^{-6}$ pc, an electron temperature of 9700$\pm$3200 K and a 
V$_{\rm{LSR}}$=24.9$\pm$2.2 km s$^{-1}$. Other radio wavelength observations include: H166$_\alpha$ RRL 
\citep{cersosimo1982}, H109$_\alpha$ and H110$_\alpha$ RRL \citep{caswell1987}, continuum emission at 5 
GHz \citep{goss1970} and 4.8 GHz \citep{wright1994}, as well as OH and H$_2$CO absorption at 1.667 and 
4.830 GHz respectively \citep{caswell1987, whiteoak1974}, $^{12}$CO(\emph{J}=1--0) emission at 115 GHz 
(\citealt{whiteoak1982, arnal2003}), and $^{13}$CO(\emph{J}=1--0) 110 GHz (\citealt{yamaguchi1999}).  
Walsh et al. (1997) attempted, unsuccessfully, to detect CH$_3$OH maser emission toward the IRAS point 
source IRAS 16362-4845. All of the RRLs and absorption features observed toward SFO 79 have very similar 
V$_{\rm{LSR}}$, located at -24$\pm$2 \kms. \citet{straw1987} conducted a detailed study at both near and 
far-infrared wavelengths of the IRAS position and the surrounding area, reporting the possible detection 
of a protostellar object and a T-Tauri star. 

All these single dish observations have large beam sizes ($>$2.8\arcmin), or are comprised of a 
single-pointing only, and therefore only allow the global properties of the molecular cloud to be
calculated and say little about the small-scale structure, or the nature of star formation, within
the BRC. We have carried out high resolution radio continuum (at 3.6 and 6 cm) and CO observations toward 
SFO 79, including the hydrogen RRL H92$_\alpha$, to study the small-scale structure of the molecular gas 
and to determine the source of the radio emission reported in Paper I. These observations have been used 
to calculate the local parameters of the molecular gas and the compact radio source. The distribution of 
emission in both sets of observations have been used to evaluate the interaction between the compact HII
region and the surrounding molecular gas. We combine our radio data with \emph{J}, \emph{H} and
\emph{K$_s$} photometric data obtained from 2 Micron All Sky Survey (2MASS; Cutri et al. 2003) and 
Midcourse Space eXperiment (MSX; Price et al. 2001) broadband mid-IR images to study the nature of star 
formation within SFO 79 and the large scale star formation in the surrounding region. 

The structure of this paper is as follows: in Section 2 we will present the details of the radio and 
molecular line observations, and a description of the archival data included in this study. Results and 
analysis are presented in Section 3. In Section 4 we present a discussion of the results and explore the 
possibility that the present star formation could have been induced. A summary of our conclusions is 
presented in Section 5.    

\section{Observations and data reduction}

\subsection{Radio observations}

Observations of SFO 79 were made using the Australia Telescope Compact Array (ATCA)\footnote{The 
Australia Telescope Compact Array is funded by the Commonwealth of Australia for operation as a National 
Facility managed by CSIRO.}, located at the Paul Wild Observatory, Narrabri, New South Wales, Australia. 
The first observation was carried out on the 22nd of September 2003 using the 750B array configuration. 
The second and third observations were carried out on the  27th and 28th of September 2003 using the 
EW357 array configuration. The ATCA consists of 6$\times$22 m antennas, 5 of which lie on a 3 km east-west
railway track with the sixth antenna located 3 km further west. This allows the antenna to be positioned 
in several configurations with a maximum baseline of 6 km. Each antenna is fitted with a dual feedhorn 
system allowing simultaneous measurements of two wavelengths, either 20/13 cm or 6/3.6 cm. The 6/3.6 cm 
receiver system was used for our observations.

The observations were made at two different frequency bands centred at 4800 and 8309 MHz (H92$_\alpha$) 
using bandwidths of 128 and 8 MHz respectively. The 8 MHz bandwidth was divided into 512 channels, giving 
a frequency resolution of 15.6 kHz and a velocity resolution of 0.56 km s$^{-1}$. SFO 79 was observed on 
each occasion for a total of one hour, split into 6$\times$10 minute snapshots spread out over a wide 
range of hour angles to optimise \emph{uv} coverage. To correct for fluctuations in the phase and 
amplitude of these data caused by atmospheric and instrumental effects, the phase calibrator 1613--586 
was observed for two minutes on either side of each snapshot observation of SFO 79. The primary flux 
calibrator 1934--638 was observed once during each set of observations to allow the absolute calibration 
of the flux density. To calibrate the bandpass the bright point source 1921--293 was also observed once 
during each set of observations. These observational parameters are summarised in 
Table~\ref{tbl:radio_parameters}.

\begin{table}[!tbp]
\begin{center}
\caption{Observational parameters for the ATCA radio observations.}
\label{tbl:radio_parameters}
\begin{minipage}{\linewidth}
\begin{tabular}{lcc}
\hline
\hline
Parameter& H92$_\alpha$$^{a}$ & 6 cm \\
\hline
Rest frequency (MHz)\dotfill & 8309 & 4800 \\
Total bandwidth (MHz)\dotfill& 8    & 128 \\
Primary beam (arcmin)\dotfill & 5 & 10 \\
Number of channels\dotfill   & 512  & 32 \\
Channel resolution (kHz)\dotfill & 15.6 & ---\\
Velocity resolution (km s$^{-1}$)\dotfill & 0.56 & ---\\
Date of observations\dotfill & \multicolumn{2}{c}{2003 Sept 22nd/27th\&28th}\\
Total integration time (hr)\dotfill & \multicolumn{2}{c}{3}\\
Phase Centre\dotfill & RA(J2000) & Dec.(J2000)\\
\dotfill		& 16$^h~40^m~00^s$ & -48$^\circ$~51\arcmin~45\arcsec \\
Array configurations \dotfill& \multicolumn{2}{c}{750B/EW357}\\
Flux density calibrator\dotfill & \multicolumn{2}{c}{1934--638}  \\
bandpass calibrator\dotfill & \multicolumn{2}{c}{1921--293}  \\
Phase calibrator\dotfill & \multicolumn{2}{c}{1613--586}  \\
\hline
\end{tabular}\\
\\
$^{a}$ We also produced the 3.6 cm continuum image from line-free channels.
\end{minipage}
\end{center}
\end{table}

The calibration and reduction of the data were performed using the MIRIAD reduction package 
\citep{sault1995} following standard ATCA procedures. The task UVLIN was used to subtract the continuum 
emission in the 8309.383 MHz bandpass from the RRL H92$_\alpha$ spectra in the \emph{uv} plane. An image
of the 3.6 cm radio continuum emission was produced from the line free channels of the H92$_\alpha$ data. 
The data were CLEANed using a robust weighting of 0.5 to obtain the same sensitivity as natural weighting, 
but with a much improved beam-shape and lower sidelobe contamination. The data obtained from baselines 
including the 6th antenna were found to distort the processed images (due to the gap in \emph{uv} 
coverage at intermediate baselines) and so were excluded from the final images. The CLEANed data were 
restored to produce images of the RRL H92$_\alpha$, 3.6 and 6 cm emission, which have synthesised 
beamwidths of $\sim$12\arcsec~and 20\arcsec~for the  H92$_\alpha$, 3.6 and 6 cm emission respectively.

\subsection{CO observations}

SFO 79 was observed during June 2003 in the \emph{J}=1--0 rotation lines of $^{12}$CO, $^{13}$CO and 
C$^{18}$O using the Mopra 22-m millimetre-wave telescope operated by the Australia Telescope 
National Facility (ATNF). The half-power beam-width at these frequencies is $\sim$30$^{\prime\prime}$ 
with a main beam efficiency of 0.4 (E. F. Ladd, private communication). Position-switching was used to 
subtract sky emission. Antenna pointing checks every two hours showed that the average pointing accuracy 
was better than 10$^{\prime\prime}$ rms. The observed antenna temperatures T$_A$ were corrected for 
atmospheric absorption, ohmic losses and rearward spillover, by taking measurements of an ambient load 
(assumed to be 290 K) placed in front of the receiver following the method of \citet{kutner1981}. The 
standard calibrators Orion KL and M17SW were observed once each day to allow for absolute flux calibration.
Hereafter all antenna temperatures quoted are on the corrected antenna temperature scale, T$_A^*$. 

The receiver is a cryogenically cooled low-noise SIS mixer with a single sideband receiver noise 
temperature of $\sim$110 K. The system temperature was found to vary by approximately 10\% due to 
weather conditions and telescope elevation. For all of our observations we used a bandwidth of 
64 MHz with a 1024-channel digital autocorrelator, giving a frequency resolution of 62.5 kHz and a 
velocity resolution of 0.16--0.17 km s$^{-1}$ over the 109--115 GHz frequency range. The 
correlator setup allows two different frequency bands to be observed. For the $^{12}$CO and $^{13}$CO 
observations the second band was tuned to 86.2 GHz (SiO maser frequency) to allow for pointing corrections 
to be performed during the observations. However, both bands were tuned to 109.782 GHz for the C$^{18}$O 
observations in order to optimise the Signal to Noise Ratio (SNR). A summary of the CO observations 
parameters are presented in Table~\ref{tbl:co_line}.

\begin{table}[!tbp]
\begin{center}
\caption{CO lines observed toward SFO 79.}
\begin{tabular}{lcccc}
\hline
\hline
Isotope  & Frequency & Resolution & Grid & Integration  \\
(\emph{J}=1--0)		& (GHz) 		& (km s$^{-1}$) & size& time (s)\\
\hline
$^{12}$CO\dotfill	& 115.271 & 0.162 		& 9$\times$9 & 30 \\
$^{13}$CO\dotfill& 110.201 & 0.170		&9$\times$9 & 30\\
C$^{18}$O\dotfill& 109.782 & 0.170		&3$\times$3  & 120  \\
\hline
\end{tabular}
\label{tbl:co_line}
\end{center}
\end{table}

The $^{12}$CO and $^{13}$CO observations consisted of a 9\x9 grid map of the cloud centred on the IRAS 
point source identified by \citet{sugitani1994}, using a grid spacing of 15\arcsec. We observed each 
point for 30 seconds interleaved with observations of an off-source reference position for 90 seconds 
after each set of 9 pointings (off-source reference position $\alpha$(J2000) =16$^h~47^m~57.1^s$, 
$\delta$(J2000)=-48$^\circ$~18$\arcmin~57\arcsec$). The C$^{18}$O maps were comprised of a smaller grid of 
3$\times$3 pointings with the same spacing, and were also centred on the IRAS position, with a total 
integration time of 2 minutes for each grid position.

The ATNF data reduction package SPC was used to process the individual spectra. Sky-subtracted spectra
were obtained by subtracting the emission from the reference position from the on-source data. A 
correction was made to account for the change in the shape of the dish as a function of elevation. 
Gaussian profiles were fitted to the individual spectra to extract the peak antenna temperature,
FWHM line widths and the velocity of the emission with respect to the local standard of rest 
(V$_{\rm{LSR}}$), listed in Table~\ref{tbl:co_data}. Maps of the integrated emission of each of the three 
lines were produced to allow us to trace the distribution of the molecular gas. The integrated emission of 
the $^{13}$CO line is presented in Figure~5, contoured over the DSS image. 

\subsection{Archival data}
	
To complement our radio and molecular line data we obtained archival data from both the 2MASS all-sky 
release atlas (\citealt{cutri2003}) and the MSX Galactic Plane Survey\footnote{The 2MASS and MSX data 
were downloaded from the NASA/IPAC Infrared Processing and Analysis Center and NASA/IPAC Infrared 
Space Archive both held at http://www.ipac.caltech.edu.} (\citealt{price2001}). The 2MASS photometric data 
were used to search the cloud for possible protostars and T-Tauri stars, also known as class I and class II 
objects respectively. The MSX data were used to study the large scale structure of RCW 108, specifically 
to identify deeply embedded thermal objects and emission from Polycyclic Aromatic Hydrocarbons (PAHs). The 
hard radiation field within HII regions destroys PAHs, which can only survive within the shielded 
environment of molecular clouds on the edges of HII regions. UV photons penetrate the surface of molecular 
clouds, leading to the formation of a Photo Dominated Region (PDR) in the surface layers of the clouds. 
PAHs within the PDR are excited by the UV photons re-emitting their energy at mid-infrared wavelengths, 
particularly between 6-10~$\mu$m. For this reason the MSX 8 $\mu$m images are excellent tracers of PDRs, 
which themselves indicate the presence of a nearby high-mass star (Rathborne et al. 2002).

We obtained photometric measurements from the 2MASS Point Source Catalogue (\citealt{cutri2003}) in the 
\emph{J}(1.24 $\mu$m), \emph{H}(1.65 $\mu$m) and \emph{K$_s$}(2.16 $\mu$m) bands. The 2MASS data have a 
photometric accuracy of between 1--2\% and a positional accuracy of better than 1\arcsec. The limiting 
magnitudes are 15.8, 15.1 and 14.3 for \emph{J}, \emph{H} and \emph{K$_s$} respectively. In order to 
concentrate on star formation within the BRC we obtained photometric data on a total of 34 sources from 
the 2MASS catalogue which lay within a 90\arcsec~radius of the position of the IRAS point source. We 
excluded any 2MASS sources that had a SNR less than 10 in any of the three wavebands.

We obtained one degree field images of RCW 108 in all four of the MSX wavebands, band A (8.3 $\mu$m), 
band C (12.1 $\mu$m), band D (14.7 $\mu$m) and band E (21.3 $\mu$m). Each MSX band covers a discrete 
bandpass of between 2--6 $\mu$m wide, here we have only quoted the centre wavelength. Together these 
four images provide information about the large scale interaction between the OB stars and the 
surrounding molecular gas. The most useful wavebands for this study were band A and band E. The 
bandpass of band A includes several discrete PAH emission features (e.g. 7.7 and 8.6 $\mu$m) 
as well as a thermal continuum component from hot dust at $\sim$400 K. The emission in band E arises due to warm dust at $\sim$100 K 
and is very useful for tracing warm dust and identifying deeply embedded objects such as YSOs and 
compact HII regions. Where the band A and band E emission are spatially well correlated we have assumed
that the emission is dominated by warm dust, thus suggesting an embedded source. Conversely, where the 
emission in these bands is not spatially correlated, we have assumed that it is dominated by PAH emission.

\section{Results and analysis}
\subsection{H92$_\alpha$ radio recombination line}

\begin{figure}[!tbp]
\begin{center}
\includegraphics[scale=0.58]{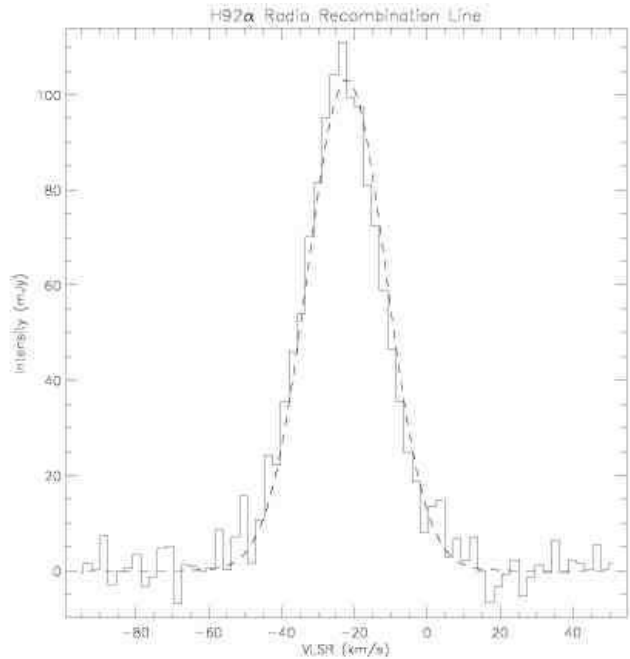}
\caption{The source averaged H92$_\alpha$ spectrum (histogram) fitted with a dashed line indicating 
the single Gaussian profile fitted to the data.}   
\label{fig:rrl}
\end{center}
\end{figure}

We detected bright emission from the H92$_\alpha$ RRL toward IRAS 16362--4845, confirming the 
presence of an embedded compact HII region within SFO 79. The RRL was Hanning smoothed to improve the SNR, 
reducing the velocity resolution to $\sim$1 km s$^{-1}$. A single Gaussian profile was fitted 
to a source averaged spectrum of the data using a least-squares fitting routine to obtain the line 
parameters (see Figure~\ref{fig:rrl}). The central velocity of the line (V$_{\rm{LSR}}$), the FWHM 
line width ($\Delta$V), and the peak line emission were found to be -24 km s$^{-1}$, 24.6  km s$^{-1}$ 
and 104 mJy respectively. The V$_{\rm{LSR}}$ of the line compares very well with the value of -23 
km s$^{-1}$ determined from H$_\alpha$ observations by \citet{georgelin1996} and we will therefore adopt 
their distance estimate to the HII region RCW 108 of 1.35 kpc as the distance to SFO 79.  The line width 
compares well with values typically observed toward UC HII regions which range between 25-30 
\kms~\citep{wood1989}.    

\begin{figure*}[!tbp]
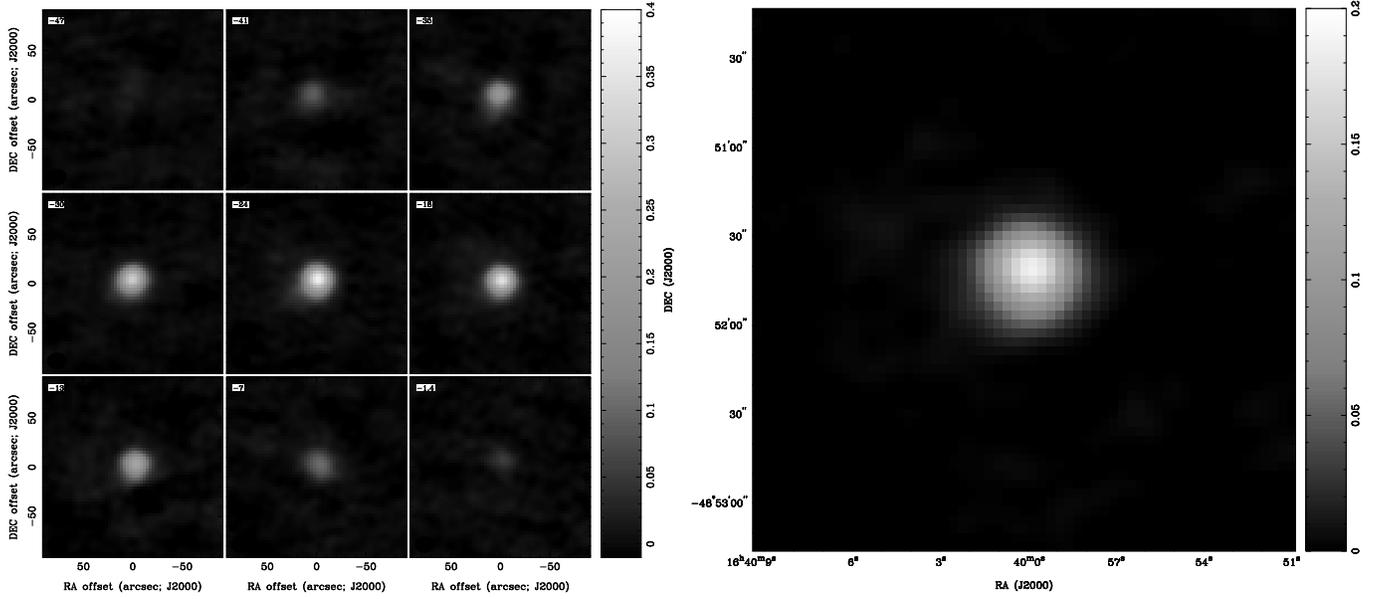

\begin{center}
\includegraphics[scale=0.40, angle=-90]{485fig3a.ps}
\includegraphics[scale=0.40, angle=-90]{485fig3b.ps}\\
\caption{The images above display the results of the data reduction on the RRL data. \emph{Left 
panel:} Nine velocity slices through the spectral data cube in steps of 5.6 km s$^{-1}$. Each slice is 3.4 km 
s$^{-1}$ wide. The velocity distribution suggests the compact HII region has a spherical morphology. 
\emph{Right panel:} Image of the integrated velocity channels, integrated between the V$_{LSR}$ range 
-50 to 0 km s$^{-1}$.}

\label{fig:velocity_slices}
\end{center}
\end{figure*}

The line-to-continuum ratio, \emph{$T_L/T_C$}, was calculated by dividing a source summed spectrum by the 
source integrated 3.6 cm continuum emission and was found to be 0.109. The RRL can be used to derive the 
electron temperature (\emph{$T_e$}) of the radio emission from the \emph{$T_L/T_C$} and linewidth. We used 
Eq. 1 from \citet{garay1994} to calculate \emph{$T_e$}, making the following assumptions: the average 
ionised helium to hydrogen abundance (\emph{Y$^+=He^+/H^+$}) for our galaxy is 0.074$\pm$0.003 
(\citealt{shaver1983}); the emission from both the RRL and the continuum is optically thin; and the 
emission is thermal in origin and emitted under Local Thermodynamical Equilibrium (LTE) conditions. We 
determined \emph{$T_e$} to be 7700 K. A summary of the derived parameters and their associated errors can 
be found in Table~\ref{tbl:recombination_line}. (The quoted errors are 1$\sigma$ errors associated with 
the Gaussian fit only). Figure~\ref{fig:velocity_slices} shows an integrated velocity image  
(-50 to 0 \kms) and velocity channel maps of the RRL emission. The channel maps show the HII region to 
have spherical morphology at all velocities with the high velocity components lying in a direction 
parallel to our line of sight.

\begin{table}[!tbp]
\begin{center}
\caption{Gaussian properties of the source averaged H92$\alpha$ radio recombination line.}
%\smallskip
\begin{tabular}{lc}
\hline
\hline
H92$_\alpha$ Parameter & Measurement \\
\hline
T$_l$/T$_c$\dotfill & 0.109$\pm$0.004 \\
V$_{\rm{LSR}}$ (km s$^{-1}$)\dotfill  & -24$\pm$0.3 \\
$\Delta$V (km s$^{-1}$)\dotfill  & 24.6$\pm$2.4 \\
T$_e$ (K)\dotfill  & 7700$\pm$1200  \\
\hline
\end{tabular}
\label{tbl:recombination_line}
\end{center}
\end{table}

\subsection{Radio continuum emission}

\begin{figure*}[!tbp]
\begin{center}
\includegraphics[scale=0.85]{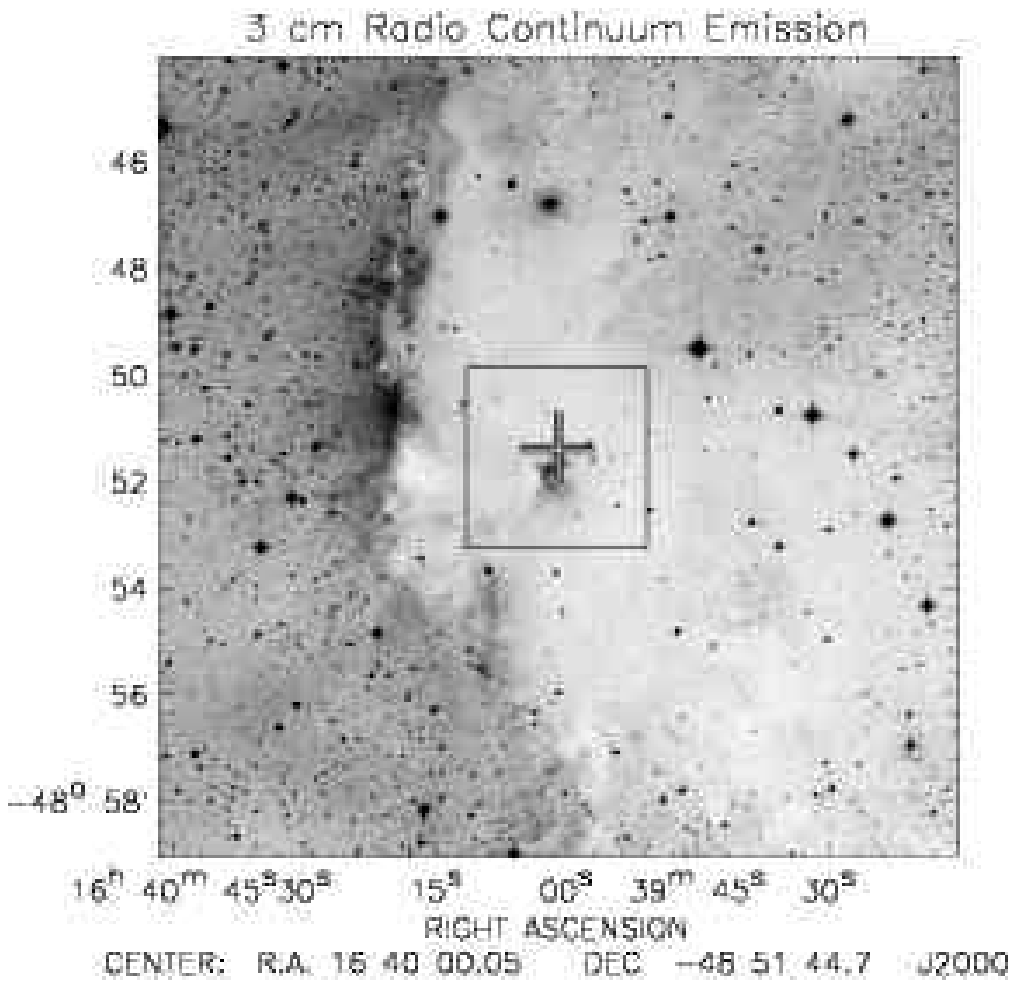}
\includegraphics[scale=0.85]{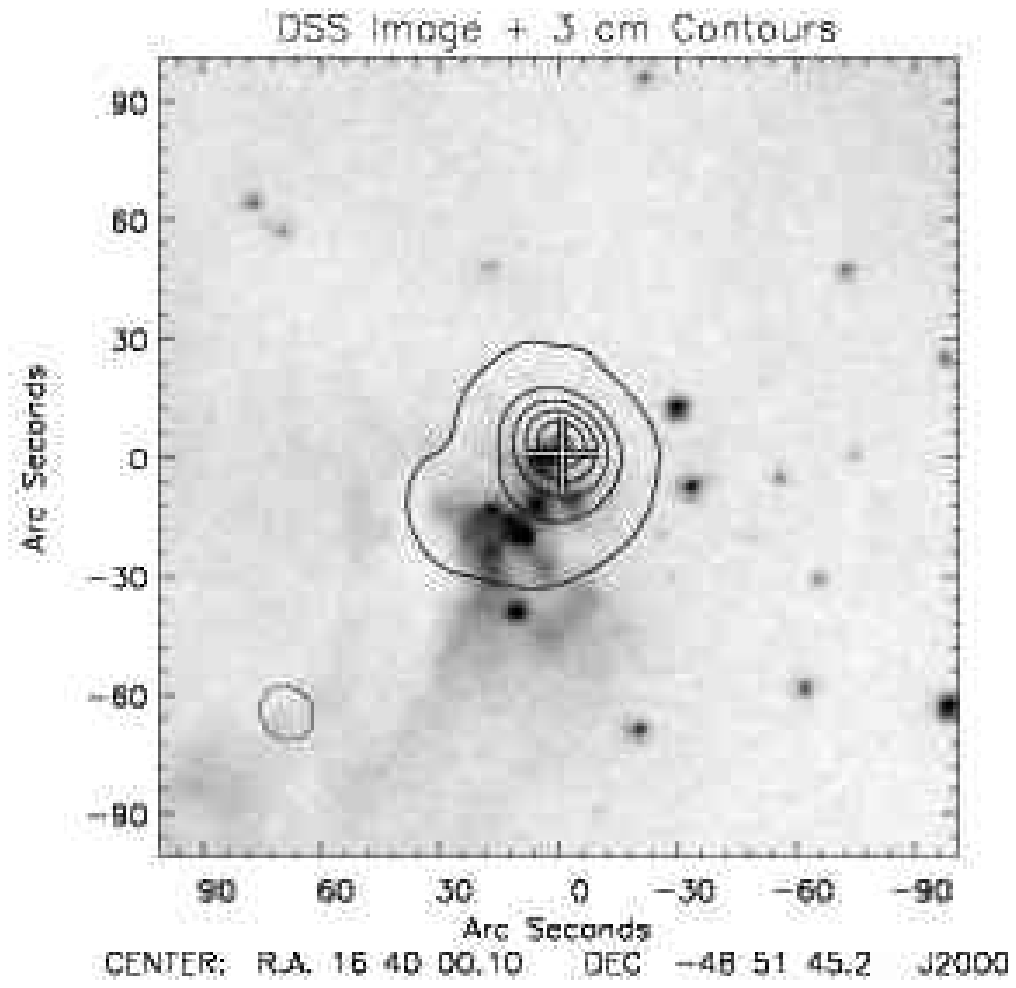}\\
\includegraphics[scale=0.85]{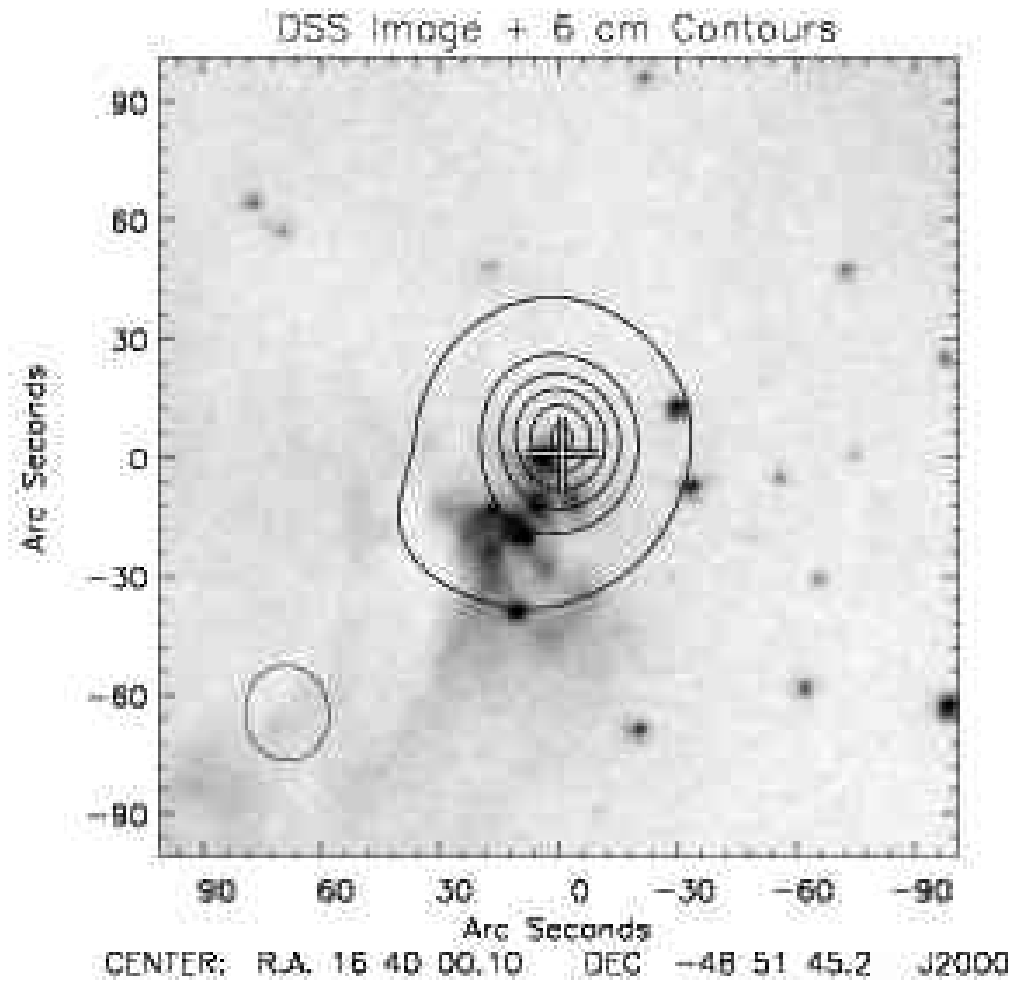}
\includegraphics[scale=0.85]{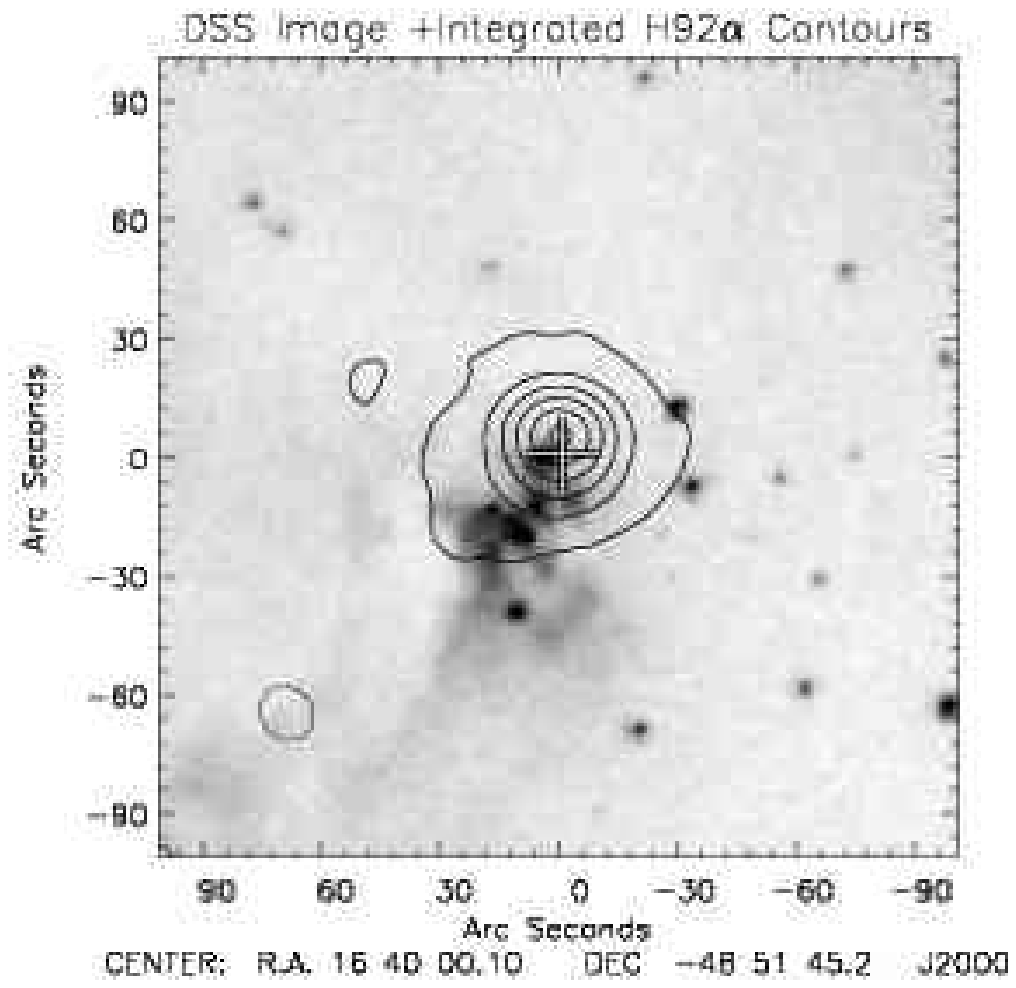}\\
\caption{DSS image  (R band) of SFO 79 and results of the radio observations. \emph{Top left panel}: 
Large scale DSS image of SFO 79 centred on the position of the IRAS point source. The central cross 
indicates the position of the IRAS point source, the black outlined square indicates the area displayed 
in closeup in the \emph{Top right} and \emph{bottom panels}. \emph{Top right and lower left panels}: 
Contours of the 3.6  and 6 cm radio continuum emission overlaid on a closeup of the DSS image of SFO 79. 
\emph{Lower right panel}: Integrated H92$_\alpha$ emission contours overlaid on the DSS image. Contours 
start at $\sim$3$\sigma$ and increase in six equal fractions of the peak flux. The synthesised beams for 
each wavelength are outlined in black in the lower left corner of each image.}   

\label{fig:radio}
\end{center}
\end{figure*}

The radio continuum and integrated H92$_\alpha$ emission images are presented in Figure~\ref{fig:radio} 
and the measurements of the peak and source integrated emission are summarised in Table~\ref{tbl:image}. 
The image analysis has been performed using the visualisation package \emph{kvis}, which is part of the 
\emph{Karma} software suite \citep{gooch1996}. The radio emission is very similar in each of the three 
close-up images in Fig.~\ref{fig:radio}, although one difference can be seen between the continuum 
emission and the RRL emission; the emission distribution shown by the 3.6 and 6 cm continuum contours 
would suggest the compact HII region to have a core-halo morphology, however, this is not observed in the 
RRL emission, which would suggest a spherical morphology. This disagreement is most likely a sensitivity 
effect, as both of the continuum images are considerably more sensitive than the RRL observations and thus 
more likely to identify fainter emission from the halo (i.e. continuum map rms $\sim$4-9 mJy compared to 
RRL image rms $\sim$77 mJy). We consider that the higher sensitivity continuum images provide the most 
accurate means of determining the morphology of the HII region, and therefore, we have adopted a core-halo 
morphology for the HII region for the remainder of this paper. 

The positions of the emission peaks at all three wavelengths agree with each other to within a few 
arc-seconds, and correlate well with the position of the IRAS point source, 16362--4845, which is offset 
to the south by $\sim$7\arcsec. Taking into account the IRAS beam size (0.5\arcmin--2\arcmin~between 12 
and 100 $\mu$m), we consider these to be in excellent agreement. Comparing the integrated flux with the 
flux measured by the Parkes-MIT-NRAO (PMN; source id. PMNJ1640-4851) survey \citep{wright1994} of $\sim$8.5 
Jy (4.850 GHz) and that reported by \citet{caswell1987} of $\sim$7.6 Jy, and because the spectrum is flat 
between 4.8 and 8.3 GHz ($\sim$ -0.03, see next paragraph), we find that our observations miss 
$\sim$12--25\% of the total flux density. This is because the Compact Array is an interferometer and as 
such works as a high spatial frequency filter, filtering out emission from large scale structure. However, 
we note that this difference in the flux density does not affect our measurement of the line to continuum 
ratio because both the line and continuum measurements are equally affected.

The spectral index ($\alpha$) is defined by the relationship $S_\nu \propto \nu^\alpha$, where  $S_\nu$ 
is the flux density of the source and $\nu$ is the frequency at which the emission is measured. For a 
thermal plasma we would expect $\alpha$ to range from -0.1 to 2 depending on the optical thickness of 
the source, where -0.1 would infer that the plasma was optically thin \citep{mehringer1994}. Using the
3.6 and 6 cm source integrated fluxes we have calculated the spectral index of the radio emission to 
be -0.027, therefore the emission is both thermal and optically thin.   
   
\begin{table*}[!tbp]
\caption{Summary of physical parameters derived from radio observations.}
\smallskip
\begin{center}
\begin{tabular}{lccccccc}
\hline
\hline
	Image		 &  FWHM restoring 	&Position 	&Peak   		& Integrated  	& Deconvolved Source & Physical 	&Image  \\
				 &	       beam 		&angle 		& Emission 	& Emission 		& FWHM 	& size		& rms noise\\
			& (arcsec)				& (degrees)	&(Jy/beam) 	& (Jy) 			& (arcsec) & (pc)		&(mJy/beam) \\
\hline
H92$_\alpha$ &12.7$\times$14.3 & 30.6  &0.18 &0.24& 11.2& 0.07 & 77\\
3.6 cm   & 12.7$\times$14.3 & 38.7 &   2.6 &6.8& 16.6& 0.11 & 4.5 \\
6 cm &  20.5$\times$24.0 & 6.4 &4.3 & 6.9 & 17.5 & 12 & 8.8 \\
\hline
\end{tabular}
\label{tbl:image}
\end{center}
\end{table*}

To estimate the size of the emission regions at different frequencies we fit Gaussian profiles to the 
azimuthal flux averages of the three images, taking into account the beam sizes by assuming a simple convolution 
of our Gaussian source with a Gaussian beam, i.e., 
$\Theta_{\rm{source}}$=$\sqrt{\Theta^2_{\rm{obs}}-\Theta^2_{\rm{beam}}}$ where $\Theta$ is the FWHM. The 
results of this analysis are also tabulated in Table~\ref{tbl:image}. These angular diameters were used 
to calculate the physical diameter of the compact HII region, by assuming that the distance to RCW 108 
and the compact HII region are the same (1.35 kpc). By taking the average of the observations at each
wavelength we obtain an effective physical diameter \emph{D$_{\rm{eff}}$} of $\sim$0.1 pc, which is 
typical for an UC HII region \citep{wood1989}. UC HII regions are associated with bright radio and 
infrared sources that appear to be deeply embedded within their natal molecular clouds and therefore are 
thought to represent an early stage in the evolution of massive stars. 

The equations of \citet{mezger1967} and \citet{panagia1978} have been used to calculate the radio 
parameters of the UC HII region, namely the Emission Measure (EM), the electron density (\emph{n$_e$}) and 
the total mass of the ionised gas (\emph{M$_{\rm{HII}}$}). The total ionising photon flux (\emph{N$_i$}) 
has been calculated using the modified version of Eqn 7 from \citet{carpenter1990} presented in Paper I;

\begin{equation}
N_i=7.7\times10^{43}S_{\nu}D^2\nu^{0.1}
\label{eq:total_flux}
\end{equation}

\noindent where \emph{N$_i$} is the total number of photons emitted per second, \emph{S}$_\nu$ is the 
integrated radio flux in mJy, \emph{D} is the distance to the source in kpc and $\nu$ is the frequency
of the observation in GHz. The results of these calculations for the two wavelengths were averaged and 
are presented in Table~\ref{tbl:radio_cal}. By assuming that the expansion speed of the UC HII region is 
the same as the sound speed in an ionised gas (i.e., $\sim$11.4 km s$^{-1}$  assuming an initial density 
of 10$^2$ cm$^{-3}$ for the surrounding material (\citealt{thompson2004b}), we estimate a lower limit of 
$\sim$10$^4$ yrs for the age of the UC HII region. The following assumptions have been made when 
calculating the UC HII region's physical parameters: the HII region is ionisationally bounded, spherical, 
has a uniform density and UV absorption by dust grains is negligible. The combined effect of the core-halo 
morphology and absorption of UV by dust on our assumptions would lead to an underestimation of the value 
of \emph{N$_i$} determined from the integrated flux. Therefore our calculated value for \emph{N$_i$} 
should be considered as a lower limit.

\begin{table}[!tbp]
\caption{Summary of ultra compact HII region parameters determined from the radio continuum data 
(see text for details).}
\smallskip
\begin{center}
\begin{tabular}{ll}
\hline
\hline
Parameter			 &  Measurement\\ 	
\hline
\emph{M$_{\rm{HII}}$} (M$_\odot$)\dotfill & 0.22 \\
\emph{n$_e$} (cm$^{-3}$)\dotfill & 13800  \\
%\emph{$\tau$}\dotfill & 0.14 \\
\emph{\rm{Log}(N$_i$})  \dotfill & 48.06 \\ 
\emph{EM} (cm$^{-6}$ pc)\dotfill & 9.9$\times$10$^6$\\

\hline
\end{tabular}
\label{tbl:radio_cal}
\end{center}
\end{table}

Assuming that the radio emission observed toward the embedded IRAS point source is produced by a single 
ZAMS star, we estimate the spectral type of the central star by comparing the derived value of the total 
number of ionising photons to values tabulated by \citet{panagia1973}. From this comparison we estimate 
the spectral type of the star exciting the UC HII region to be O8.5--O9. This is slightly earlier than 
the spectral type suggested in Paper I due to the improved \emph{uv} coverage and higher resolution of 
our observations. We find that our radio data inferred the presence of an O8.5--O9 star, which is three 
spectral classes later than the O6--O6.5 ZAMS star calculated by integrating the flux under an assumed 
blackbody curve fitted to the far-IR data (\citealt{straw1987}). The difference between these two values 
may indicate the presence of a cluster of low to intermediate mass stars that are contributing to the 
infrared luminosity, or that a significant fraction of the UV radiation is being absorbed by the dust. 
\citet{wood1989} showed that for a realistic initial mass function the spectral type of the most massive 
member of a stellar cluster would be approximately 1.5--2 spectral classes lower than that calculated 
assuming the far-IR flux was from a single star. Therefore the difference between the spectral types 
calculated from the far-IR and radio fluxes is consistent with emission expected from a stellar cluster.

\subsection{Molecular line data} 

\citet{arnal2003} surveyed the whole region in $^{12}$CO(\emph{J}=1--0). Their survey covered an area of 
2.75$^\circ$$\times$3.00$^\circ$ with an angular resolution of 8.7\arcmin~(FWHM) and mapped the 
distribution of molecular material. The material associated with Ara OB1 was found to span a velocity 
range between -28 to -20 \kms. The most negative radial velocities was found toward the interface between 
the RCW108 HII region and the dense material to the west of the HII region (see Figure~\ref{fig:rcw108}). 

The $^{12}$CO emission revealed a large molecular condensation toward SFO 79. The condensation is located 
at $\alpha$(J2000)=$16^h~39^m~55^s$, $\delta$(J2000)=\mbox{-48$^\circ$~55\arcmin~2.1\arcsec}~which has a 
V$_{\rm{LSR}}$ of 23.5 \kms, an excitation temperature of 9.5 K. \citet{arnal2003} calculated the size of 
the condensation to be 4.7 pc$\times$2.7 pc with a mass of $\sim$3900 M$_\odot$ and a particle density of 
4.4$\times10^{3}$~cm$^{-3}$. \citet{arnal2003} concluded that the 5 GHz radio continuum emission detected 
by \citet{goss1970} and the bright IRAS point source 16362-4845 were both observational manifestations of 
an on-going star formation process within the molecular condensation and that the positional coincidence 
of other IRAS point sources with other nearby $^{12}$CO condensations may suggest that star formation is 
widespread throughout the whole molecular complex. We have made CO maps of a 2\arcmin$\times$2\arcmin~region
within the molecular condensation identified by  \citet{arnal2003} in order to examine the star formation 
indicated by the presence of the IRAS point source and the radio continuum emission presented in Section 3.2.

\begin{figure}[!tbp]
\begin{center}
\includegraphics[scale=0.5]{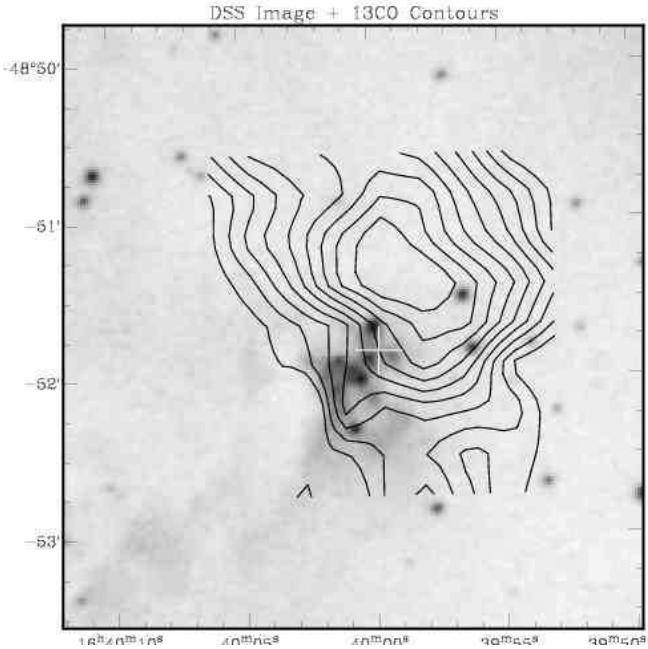}
\caption{DSS images overlaid with contours of the integrated $^{13}$CO emission. The contours start at 
30\% of the peak emission, increasing in 7\% intervals. The image reveals the presence of a molecular 
condensation offset from the position of the IRAS point source.}
\label{fig:pointings}
\end{center}
\end{figure}

\begin{figure*}
\begin{center}
\includegraphics[scale=0.95]{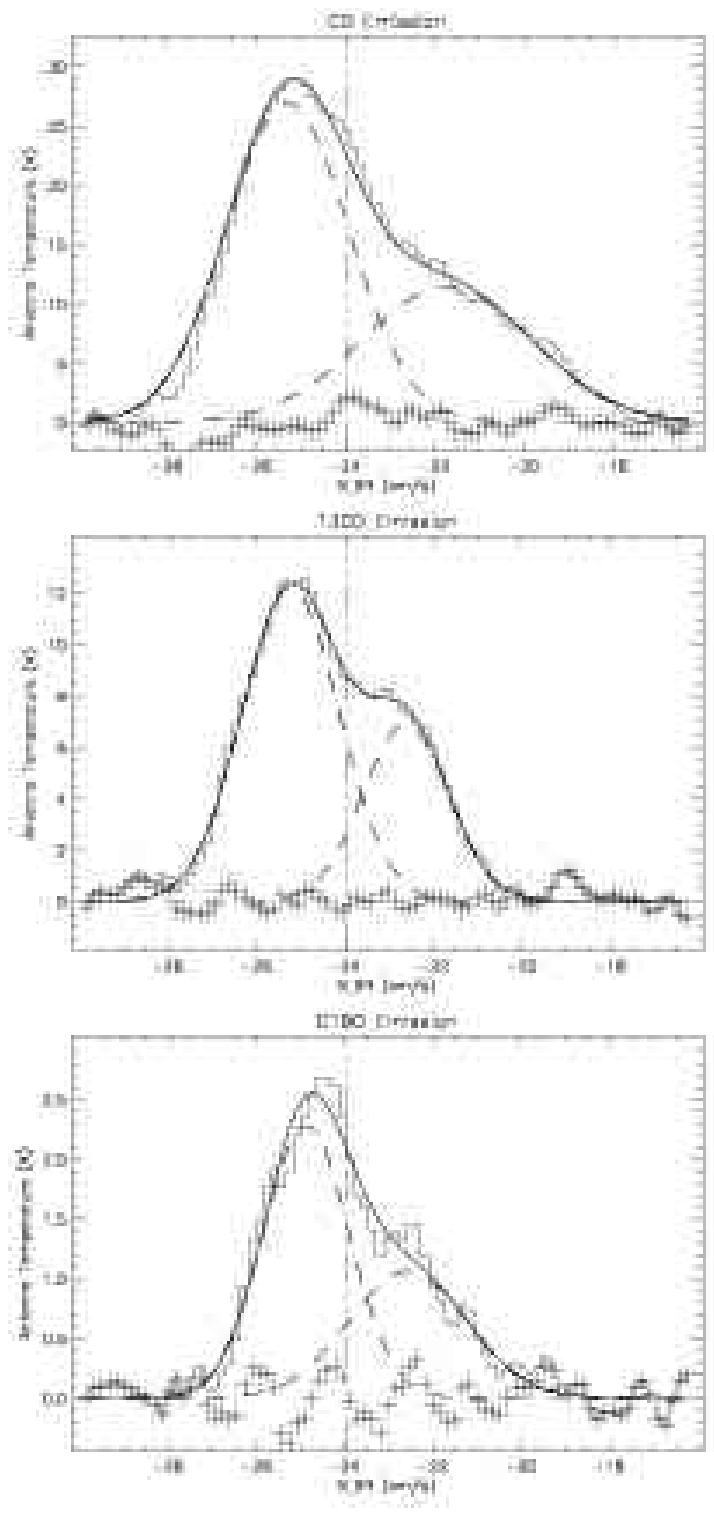}
\includegraphics[scale=0.95]{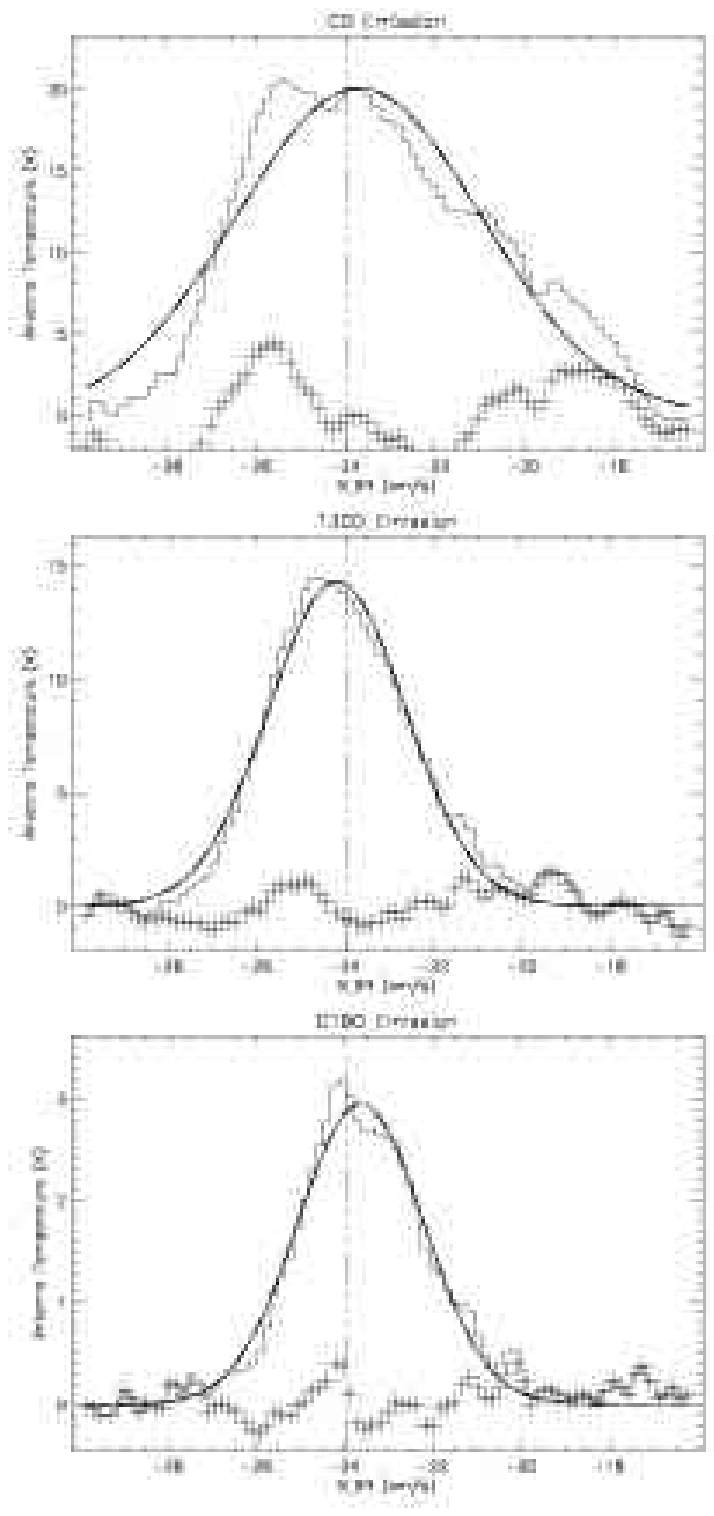}
\caption{Molecular line emission detected toward SFO 79. \emph{Left panel}: emission detected toward the 
IRAS point source and the peak of the radio detection. \emph{Right Panel}: emission from the molecular 
core detected to the northwest of the IRAS point source. The observed data are shown as a histogram with 
the Gaussian fit to the data represented by a solid line. Where a combination of two Gaussian profiles to 
the data have been used, the individual components are shown by  dashed lines. Residuals are indicated by 
the crosses. The dashed-dotted vertical line indicates the velocity of the HII region derived from the 
radio observations.}
\label{fig:co_spectra}
\end{center}
\end{figure*}

Figure~\ref{fig:pointings} shows the integrated emission map of $^{13}$CO superimposed over a DSS image of 
SFO 79. The contoured image reveals the presence of a molecular condensation 30\arcsec~to the northwest of 
the IRAS position. We present spectra of the CO lines observed toward the molecular condensation and the 
IRAS point source in Figure~\ref{fig:co_spectra}. The vertical line in Fig.~\ref{fig:co_spectra} represents 
the central line velocity of the H92$_\alpha$ line observed toward the UC HII region. The spectra of all 
three isotopomers taken close to the position of the IRAS point source suggest the presence of two 
velocity components, at approximately -22 km s$^{-1}$ and -25 km s$^{-1}$ (see Figure~\ref{fig:co_spectra}).
These velocity components lie on either side of the velocity of the UC HII region as measured from the 
H92$\alpha$ line (-24 km s$^{-1}$). These CO spectra were found to be best fitted with a combination of 
two Gaussian profiles. The detection of two molecular components could be due to the expansion of the 
HII, the molecular gas on either side of the HII region being accelerated both toward and away from us 
along our line of sight. The spectra observed toward the molecular condensation are comprised of a single 
velocity component that has a peak velocity of $\sim$24 km s$^{-1}$, correlating extremely well with the 
central velocity of the UC HII region. The peak antenna temperature (\emph{T$_A^*$}), V$_{\rm{LSR}}$ and 
the FWHM line width are presented in Table~\ref{tbl:co_data}.
%The differences in the line intensities and width could be 
%explained by an asymmetric expansion of the HII region into regions of different molecular densities. 

The $^{12}$CO and $^{13}$CO data display some evidence of moderate velocity non-Gaussian line wings (see 
Figure~\ref{fig:co_spectra}) indicative of molecular outflows over the 2\arcmin\ field, however, outflow 
lobes were not observed in the channel maps or integrated intensity maps of either line. This is probably 
due to the insufficient resolution of the observations. Higher resolution molecular line observations are 
needed to obtain a clearer picture of the small scale structure of the molecular condensation and to 
detect the presence of a molecular outflow.

\begin{table}
\caption{Results of the Gaussian fitting of the CO spectra observed toward the IRAS point source and the 
molecular condensation offset to the northwest.}
\begin{center}
%\scriptsize
\begin{tabular}{lccc}
\hline
\hline
Molecular   & V$_{LSR}$  & FWHM & Peak T$_A^*$   \\
Line & (kms$^{-1}$) & (kms$^{-1}$) & (K) \\
\hline
\multicolumn{4}{c}{IRAS point source}\\
\hline
%CO\dotfill& -25.3$\pm$0.0005   &  1.3$\pm$0.0004  &  25.2$\pm$0.202\\
$^{12}$CO\dotfill& -25.3   &  1.3  &  25.2\\
%\dotfill& -22.1$\pm$0.0056   &  2.1$\pm$0.0031  &  12.1$\pm$0.046\\
\dotfill& -22.1   &  2.1  &  12.1\\

%$^{13}$CO\dotfill & -25.2$\pm$0.0017   &  1.1$\pm$0.0012  &  12.3$\pm$ 0.382\\
$^{13}$CO\dotfill & -25.2   &  1.1  &  12.3\\

%\dotfill				&-22.6$\pm$0.1372    & 1.0$\pm$0.0466  &  6.9$\pm$0.169\\
\dotfill				&-22.6 & 1.0  &  6.9\\
%C$^{18}$O\dotfill & -24.8$\pm$0.0039  &   1.0$\pm$0.0019  &   2.3$\pm$ 0.0111\\
C$^{18}$O\dotfill & -24.8  &   1.0  &   2.3\\

%\dotfill&			 -22.6$\pm$0.0316 &    1.4$\pm$0.0159  &   1.1$\pm$ 0.0062\\
\dotfill&			 -22.6 &    1.4  &   1.1\\
\hline
\multicolumn{4}{c}{Molecular Condensation}\\
\hline
%CO\dotfill& -23.7$\pm$0.0003   &  2.8$\pm$0.0002  &  20.1$\pm$0.8590\\
$^{12}$CO\dotfill& -23.7   &  2.8  &  20.1\\

%$^{13}$CO\dotfill &-24.2$\pm$0.0002   &  1.6$\pm$0.0002  &  14.3$\pm$0.109\\
$^{13}$CO\dotfill &-24.2  &  1.6  &  14.3\\
%C$^{18}$O\dotfill &-23.6$\pm$ 0.0016  &   1.4$\pm$0.0013  &   3.0$\pm$0.017\\
C$^{18}$O\dotfill &-23.6  &   1.4  &   3.0\\
\hline
\end{tabular}
\label{tbl:co_data}
\end{center}
\end{table}

We have used the optically thin transition of C$^{18}$O and the moderately optically thick ($<$1) 
$^{13}$CO to determine the physical parameters of the molecular condensation. Unlike the $^{12}$CO 
transition, which is optically thick and therefore probes the conditions at the surface of unit optical 
depth, the C$^{18}$O observations are optically thin, and are thus able to sample the molecular gas 
through the whole cloud. These lines are also free of self absorption and beam filling factors. The 
optical depth can be calculated numerically using the relationship between brightness temperature and 
optical depth,
 
%\[\left(\frac{T_{13}}{T_{18}}\right)=\left(\frac{1-e^{-\tau_{13}}}{1-e^{-\tau_{18}}}\right)\]
\begin{equation}\label{eq:optical_depth}
\frac{T_{13}}{T_{18}}=\frac{1-e^{-\tau_{13}}}{1-e^{-\tau_{18}}}
\end{equation}

\noindent where \emph{T$_{13}$}, \emph{T$_{18}$} are the corrected antenna temperatures and 
\emph{$\tau_{13}$}, \emph{$\tau_{18}$} are the optical depths of the $^{13}$CO and C$^{18}$O transition 
lines respectively. The optical depths are related to each other by their abundance ratio 
such that \emph{$\tau_{13}$}=X\emph{$\tau_{18}$}. We have assumed the abundance ratio between $^{13}$CO
and C$^{18}$O to be $\sim$ 10 (\citealt{tothill2002}), that the molecular condensation is in LTE (and 
therefore the excitation temperature is the same for both transitions) and that the beam filling 
factors are similar.

Following the method described in \citet{myers1983}, we estimated the excitation temperature 
\emph{T$_{ex}$} $\sim$20 K and the column density N(C$^{18}$O) $\sim$5.5$\times$10$^{15}$ cm$^{-2}$. 
Assuming a fractional abundance of (C$^{18}$O/H$_2$)=1.7$\times$10$^{-7}$ \citep{goldsmith1996} we 
estimate the H$_2$ column density to be N$(H_2)$=3.3$\times$10$^{22}$ cm$^{-2}$. The angular size of 
the molecular condensation was calculated from the FWHM of the azimuthally averaged $^{13}$CO integrated 
intensity map (see Figure~\ref{fig:pointings}) and was found to be 1\arcmin, which corresponds to a 
physical diameter of 0.4 pc at a distance of 1.35 kpc. Assuming the condensation to be spherical, and 
that emission from other sources along the line of sight can be neglected, we find the volume density to 
be n$_{H_2}$=2.7$\times 10^4$~cm$^{-3}$. Multiplying this by the total volume of the condensation gives 
an upper limit for the mass. The following equation has been used to calculate the total mass of the 
condensation in units of M$_\odot$,

\begin{equation}\label{eq:mass}
M_{\rm{Cond}}=6.187\times10^{25}R^3n_{\rm{H_2}}\mu m_{\rm{H}},
\end{equation}

\noindent where \emph{R} is the radius of the condensation in parsecs, \emph{n$_{\rm{H_2}}$} is the 
molecular hydrogen volume density in cm$^{-3}$, $\mu$~is the mean molecular weight (taken to be 2.3, 
assuming a 25\% abundance of helium by mass), and \emph{m$_{\rm{H}}$} is the atomic mass unit. The mass 
was calculated to be $\sim$51 M$_\odot$, which is rather low for a high mass star forming region. However, 
this could be due to depletion of the CO molecules onto either dust grain ice mantles or via selective 
photo-dissociation. Comparing the core mass to those derived from sub-millimetre and CO observations of 
similar clouds \citet{thompson2004b} found that, on average, masses obtained using CO data gave core 
masses roughly an order of magnitude lower than those produced from greybody fits to the IRAS and 
sub-millimetre data. We therefore would consider the derived mass of our condensation as a lower limit.
The physical parameters derived for the molecular condensation are summarised in Table~\ref{tbl:co_summary}.  

\begin{table}[!]
\caption{Summary of physical parameters of the molecular condensation determined from the CO data.}
\begin{center}
%\scriptsize
\begin{tabular}{cc}
\hline
\hline
Condensation's Parameters   & Measurement \\
\hline
$\tau_{18}$\dotfill & 0.163 \\ 
$\tau_{13}$\dotfill & 1.3 \\
T$_{ex}$ (K)\dotfill & 20.5\\
N(C$^{18}$O) (cm$^{-2}$)\dotfill & 5.5$\times10^{15}$ \\
N(H$_2$) (cm$^{-2}$)\dotfill & 3.3$\times10^{22}$\\
Angular diameter (arcmin)\dotfill & 1 \\
Physical diameter (pc)\dotfill & 0.4 \\
n(H$_2$) (cm$^{-3}$)\dotfill & 2.7$\times 10^4$\\
Mass (M$_\odot$)\dotfill & 51\\

\hline

\end{tabular}
\label{tbl:co_summary}
\end{center}
\end{table}

\subsection{2MASS identification of YSO candidates}

\begin{figure*} 
\begin{center}
\includegraphics[scale=0.7]{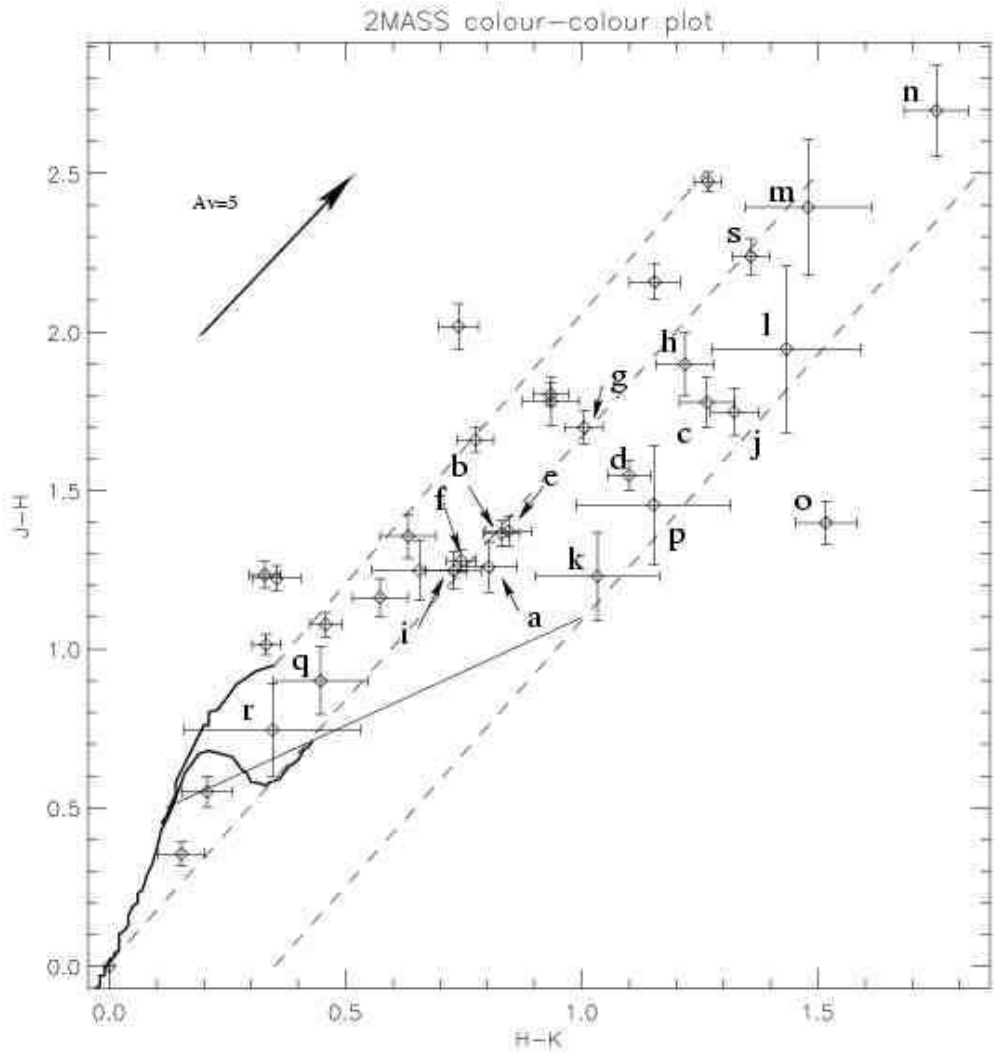}\\
\caption{\emph{J}, \emph{H} and \emph{K$_s$} colour-colour diagram of the 2MASS point sources
detected within a 90\arcsec~radius of the position of the IRAS point source. The thick solid lines 
represent the unreddened locus of both main-sequence and giant stars taken from \citet{koornneef1983}. 
The fine straight line denotes the Classical T-Tauri locus of \citet{meyer1997}. The dashed line 
indicates the reddening tracks expected due to extinction, adopted from \citet{cohen1981}. The arrow in 
the top left of the diagram indicates the effect of 5 magnitudes of extinction.}
\label{fig:jkh_image}
\end{center}
\end{figure*}

A total of 34 IR sources (SNR$>$10) were identified from a search of the 2MASS Point Source Catalogue 
(\citealt{cutri2003}) to be within a 90\arcsec~radius of the position of the IRAS point source. For each 
of these sources the \emph{J}, \emph{H} and \ks~magnitudes obtained from the 2MASS Point Source Catalogue 
were used to calculate values for their \emph{J}--\emph{H} and \emph{H}--\ks~colours. The corresponding 
colour-colour diagram, following the method of \citet{lada1992}, is presented in 
Figure~\ref{fig:jkh_image}. The photometric errors are indicated by the horizontal and vertical error 
bars. The thick solid lines represent the locus of both main-sequence and giant stars taken from 
\citet{koornneef1983}. The thin solid line represents the Classical T-Tauri Star (CTTS) locus as determined
by \citet{meyer1997}. 

\begin{figure*}[!]
\begin{center}
\includegraphics[scale=0.45]{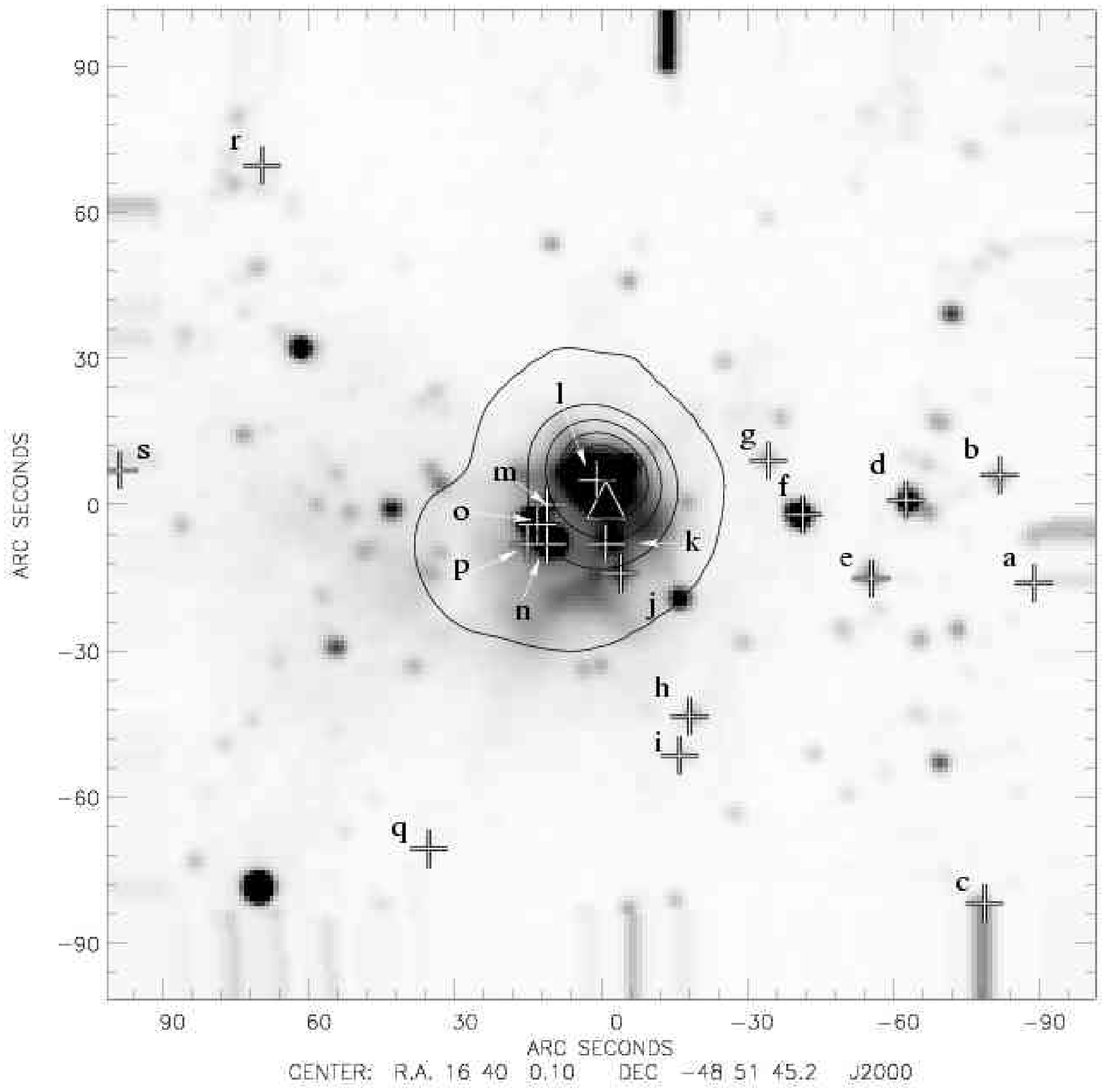}
\includegraphics[scale=0.45]{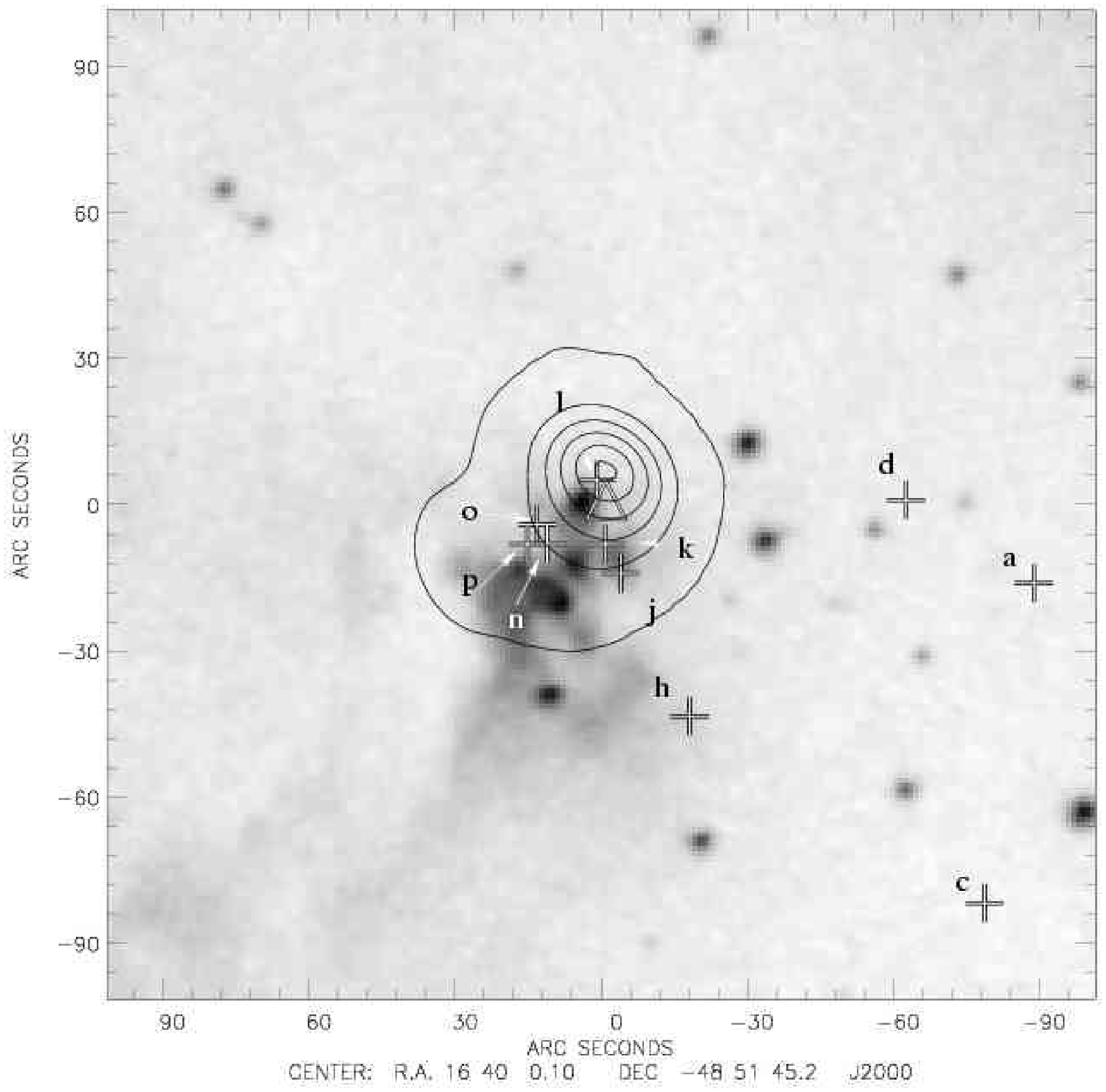}
\caption{The images above displays the candidate YSOs across a 3\arcmin~field centred on the position of 
the IRAS point source. These locations are indicated by white in black crosses, the IRAS position is 
indicated by a white in black triangle. \emph{Top panel}: 2MASS \emph{K$_s$} band Quicklook image. 
\emph{Bottom panel}: A DSS image is presented. Contours of the 3 cm continuum emission have been 
overlaid over both images to show the position and boundary of the UC HII region. The lowest contour 
starts at 3$\sigma$.}
\label{fig:protostar_location}
\end{center}
\end{figure*}

The dashed lines indicate the reddening tracks. The centre and left reddening tracks denote the boundary 
where main sequence and giant stars would be expected to lie taking account of extinction \citep{lada1992}.
The circumstellar disks surrounding CTTSs lead to the production of an excess of near-infrared emission, 
making T-Tauri stars appear more reddened than main sequence stars. Therefore sources found between the 
centre and right reddening tracks are candidate CTTSs \citep{meyer1997}. The track farthest to the right 
indicates the truncation of the T-Tauri locus of \citet{meyer1997}. We would expect sources to the right 
of this reddening track to  be protostellar candidates, reddened still further by the protostellar envelopes
surrounding the protostars. Inspection of Figure~\ref{fig:jkh_image} reveals the presence of nineteen 
CTTS/protostar candidates (hereafter candidate YSOs). The positions of these sources have been plotted onto
a 2MASS \emph{K$_s$} band Quick look image of SFO 79 and are presented in the left panel of 
Figure~\ref{fig:protostar_location} along with contours of the 3.6 cm radio continuum emission. The 
candidate CTTSs and protostar details are presented in Table~\ref{tbl:2mass_data}. 

Figure~\ref{fig:protostar_location} (left panel) reveals a fairly widespread distribution of candidate 
YSOs with the vast majority located to the west of the UC HII region. It is interesting to note the high 
concentration of candidate YSOs within the boundary of the UC HII region and that the HII region is 
extended toward them. Of the seventeen candidate YSOs identified, seven are located within the UC HII 
region indicated by the radio continuum emission contours. Taking our analysis a step further and
excluding candidate YSOs where there is some ambiguity in their classification (sources that could 
potentially be reddened main sequence stars, i.e., b, e, f, g, i, m, q, r and s) we obtain ten sources 
that we can confidently consider to be genuine YSOs. These YSOs have been plotted onto the DSS image 
presented in the right panel of Figure~\ref{fig:protostar_location}. Of the ten remaining YSOs, six are 
located within the boundary of the UC HII region. Hereafter we will refer to these candidate YSOs as an 
IR cluster. The IR cluster contains the only class I protostar and all of the class I protostellar 
candidates in the region, these being o, and k, l, p respectively. Moreover, analysis of the extinction 
corrected reddening values reveals that the IR cluster contains the five source with the highest infrared 
excess, which would  suggest that the IR cluster is relatively quite young, perhaps having a similar age 
to the that of the UC HII region.

\begin{table*}
\begin{center}
\caption{Summary of 2MASS sources identified as possible class I 
protostars (class~I) or classic T-Tauri stars, also known as class II protostars. The ``Source Id" refers 
to the alphabetic label given to the protostars identified in Figure~\ref{fig:protostar_location}. 
Following the classifications of Thompson et al. (2004b) we have classified these candidate YSOs as 
either class I (protostars), class II (CTT stars) or as Reddened Main Sequence (RMS) stars. Where two 
classifications have been given, the first refers to the classification from the position of the 
candidate YSOs on the H-K and J-K diagram, the second classification taking the errors into account if 
different from the first classification.}

%\scriptsize
\begin{tabular}{ccccccccc}
\hline
Source  & 2MASS  & Ra & Dec   &K$_{s}$& J - H & H - K$_{s}$ & Type  \\
Id. & PSC ID & (J2000) & (J2000) &\\
\hline
\hline
a & 16395123-4852035  &    16:39:51 &    -48:52:04  &  13.72 &  1.26$\pm$0.08 &    0.80$\pm$0.06 & class~II\\
b & 16395182-4851415  &    16:39:52 &    -48:51:41  &  12.17&  1.37$\pm$0.04 &    0.83$\pm$0.04 & class~II/RMS\\
c & 16395241-4853089  &    16:39:52 &    -48:53:09  &  12.12&  1.78$\pm$0.08 &     1.27$\pm$0.06 & class~II\\
d & 16395370-4851461  &    16:39:54 &    -48:51:46  &  10.30 &  1.55$\pm$0.05 &     1.10$\pm$0.05 & class~II\\
e & 16395449-4852014  &    16:39:54 &    -48:52:02  &  12.00&  1.37$\pm$0.05 &    0.85$\pm$0.05 & class~II/RMS\\
f  & 16395598-4851482     & 16:39:56   & -48:51:48  & 9.72 &    1.28$\pm$0.04    &0.75$\pm$0.03 & RMS/class~II\\
g & 16395662-4851366  &    16:39:57 &    -48:51:37  & 12.58&   1.70$\pm$0.05 &     1.01$\pm$0.04 & RMS/class~II\\
h & 16395828-4852287  &    16:39:58 &    -48:52:29  &  12.62&  1.90$\pm$0.10 &     1.22$\pm$0.06 & class~II\\
i & 16395848-4852368  &    16:39:59 &    -48:52:37  & 13.22 &   1.25$\pm$0.06 &    0.73$\pm$0.06 & RMS/class~II\\
j & 16400012-4851404  &    16:40:00 &    -48:51:40  & 7.74&   1.75$\pm$0.08 &     1.32$\pm$0.05 & class~II\\
k & 16395989-4851528  &    16:40:00 &    -48:51:53  & 10.48&   1.23$\pm$0.14 &     1.03$\pm$0.13 & class~II/class~I\\
l & 16395974-4851592   &   16:40:00 &    -48:51:59  & 11.42 &   1.95$\pm$0.26 &     1.43$\pm$0.16& class~II/class~I$^{a}$\\
m & 16400125-4851454  &    16:40:01 &    -48:51:45  &  10.97&  2.39$\pm$0.22 &     1.48$\pm$0.13& class~II/RMS\\
n & 16400155-4851484    &  16:40:01 &    -48:51:49  & 9.37&   2.70$\pm$0.15 &     1.75$\pm$0.07& class~II \\
o & 16400114-4851524   &   16:40:01 &    -48:51:53  &  8.79&  1.40$\pm$0.07 &     1.52$\pm$0.06& class~I$^{a}$\\
p & 16400163-4851532  &    16:40:02 &    -48:51:53  & 11.60&    1.46$\pm$0.19 &     1.15$\pm$0.16 & class~II/class~I\\
q  & 16400381-4852547     & 16:40.04    & -48:49.08 &  14.37  &0.90$\pm$0.11    & 0.45$\pm$0.10& RMS/class~II\\
r & 16400706-4850338  &    16:40:07 &    -48:50:34  & 13.87&  0.75$\pm$0.15 &    0.35$\pm$0.19 & RMS/class~II\\
s & 16401019-4851356   &   16:40:10 &    -48:51:36  & 11.76 &   2.24$\pm$0.06 &     1.36$\pm$0.04& class~II/RMS\\
\hline
\hline
\end{tabular}
\\
\label{tbl:2mass_data}
\end{center}
$^{a}$ Source l and o have been identified in \citet{straw1987} as a candidate T-Tauri and a protostar 
respectively.
\end{table*}

\subsection{Mid-infrared data}

Wide field MSX images were obtained to investigate the global star formation within the molecular cloud 
and the large scale structure of the whole region. The MSX band A image is dominated by PAH emission and 
can be used to infer the presence of a PDR. The morphology of the PDR seen toward the rim of SFO 79 is 
elongated along the length of the bright rim, and has a similar distribution to that seen in the DSS 
image, the sharp edges of both correlating extremely well, denoting the HII region/PDR boundary. 
The MSX band E image reveals the presence of three bright thermal sources embedded within the cloud, 
set back from the bright rim with respect to the ionising stars. One of these sources is coincident with 
the position of SFO 79, probably due to internal heating from an embedded main sequence star. 

\begin{figure*}[!]
\begin{center}
\includegraphics[scale=.7]{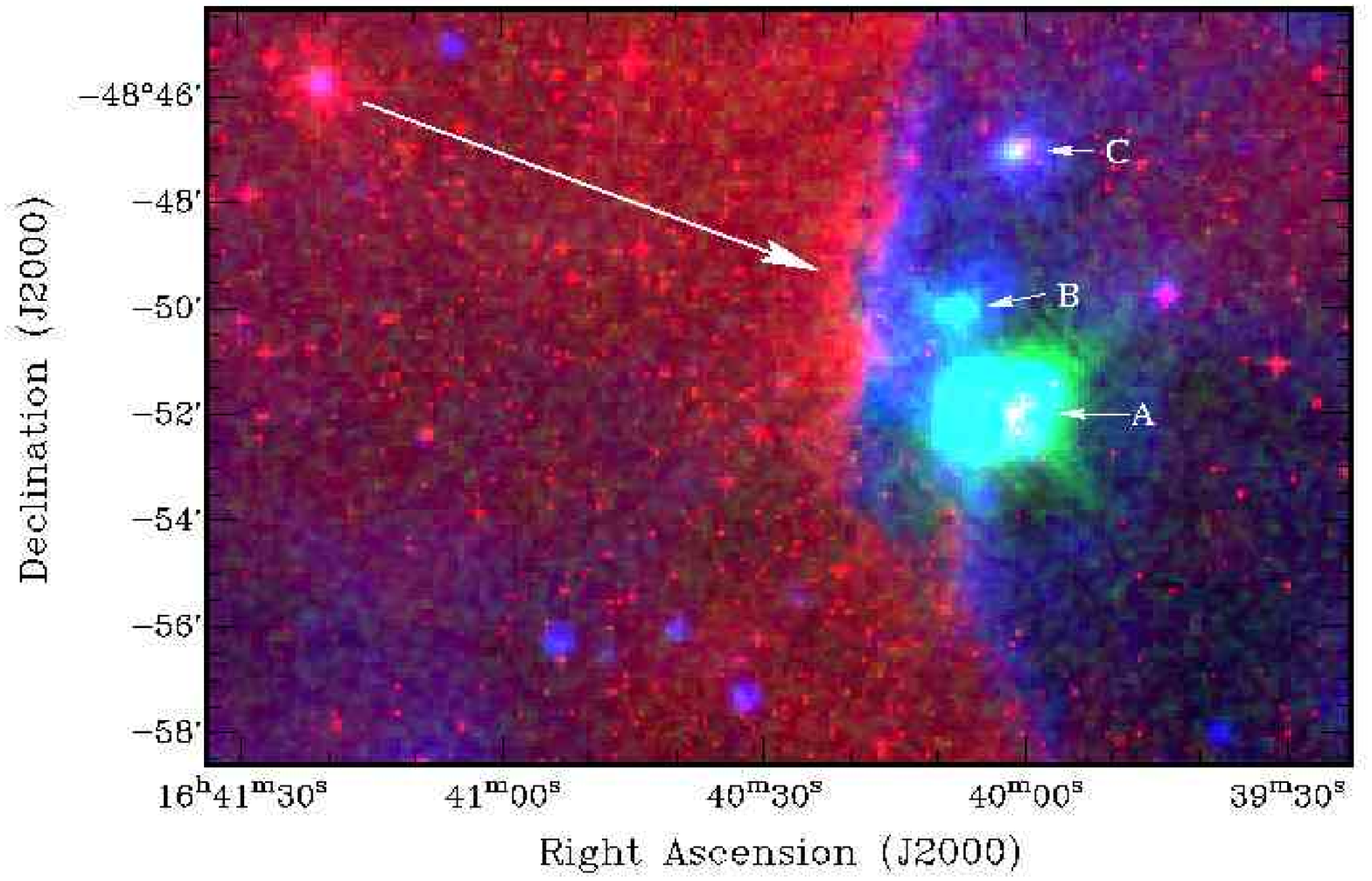}\\
\caption{\emph{Top left}: Three-colour large scale structure image of SFO 79. This image is a composite 
of the DSS image (\emph{red}), the MSX band A  (\emph{blue}) and the MSX band E  (\emph{green}). The 
band A image traces the PDRs, band E traces thermal emission due to warm dust and the DSS image traces 
the optical ionised gas. The white arrow indicates the direction of ionisation. The ionising stars can be 
seen in the upper left hand region of the image. The image clearly shows the interface between the HII 
region and the molecular material. The basic structure of the BRC shows the ionised rim followed by a PDR 
and three thermal sources labelled A, B and C which are set back from the bright rim. It is worth noting 
that all three thermal sources are at a similar projected distance from the bright rim of the cloud. 
(Colour image available in the online Journal).}
\label{fig:msx_3colour}
\end{center}
\end{figure*}

The MSX band A and E images were combined with the DSS image to produce a three colour image of the bright 
rim, the ionising stars and the surrounding region (see Figure~\ref{fig:msx_3colour}). Combining these 
images together in this way allows us to study the distribution of the mid-IR and visible wavelength 
emission and the spatial correlation between features seen in each of the three individual images. We can 
also determine the relationship between the ionised gas, the PDR, and the embedded thermal sources on a 
global scale. The image is presented in Figure~\ref{fig:msx_3colour}; band A traces emission from the 
PDRs in the region, band E traces thermal emission from warm dust, and the DSS emission traces the optically
ionised gas; these three images are coloured blue, green and red respectively. The ionising stars can be 
clearly seen as the bright red emission in the upper left section of the image. 

\section{Discussion}

In this Section we will draw together all the analyses presented in the previous Sections to investigate 
the star formation within SFO 79 as well as evaluating the global star formation in the surrounding region.

\subsection{Structure and physical properties of SFO 79}
\subsubsection{The HII/molecular cloud interface}
The interface between the HII region RCW 108 and the molecular cloud can be clearly seen in the DSS image 
(see upper left panel Figure~\ref{fig:msx_3colour}) allowing us to trace the edge of RCW 108 in great 
detail. RCW 108 is ionised by two O stars located a projected distance of $\sim$4.2 pc from the edge of 
the bright rim. Analysis of the 2MASS data for the ionising stars would suggest they are situated in a 
region of low extinction (Av$\sim$1), although the MSX band A emission which traces the PDR would suggest 
that the ionising stars are slightly in the background with respect to the cloud (see Section 4.2). No 
radio emission was detected toward the bright rim of the cloud, but the strong correlation between the 
DSS image and the MSX band A image supports the hypothesis that the surface of this cloud is being 
photoionised by the nearby O stars. Using the 6 cm image, an upper limit can be calculated for the 
impinging photon flux, which we compare to what we would expect from the ionising stars. 

The effect of the photoionisation of the cloud's surface by the OB stars is to produce a uniform 
density shell of ionised gas that surrounds the edge of the cloud known as an Ionised Boundary 
Layer (IBL). The effective thickness of this shell is $\eta$R, where $\eta$ is a geometrical factor 
of $\sim$0.2 \citep{bertoldi1989} and R is the radius of the BRC in parsecs, estimated using the 
method described in \citet{sugitani1991}. Most of the ionising radiation is absorbed in the IBL with 
the ionising photon flux being effectively balanced by radiative recombination, shielding the 
underlying molecular material. It has been shown \citep{lefloch1997} that the free-free flux measured 
in an area of width $\theta$ (FWHM) can be expressed as a function of electron density, temperature 
and source size $\theta$. In order to evaluate an upper limit to the incident ionising photon flux at the 
rim of the cloud we assume a 3$\sigma$ upper limit of 26.4 mJy at 6 cm. Rearranging Eqns 2 and 6 of 
\citet{lefloch1997}, the following two equations for the total incident flux, $\Phi$ (photons 
cm$^{-2}~s^{-1}$), and the electron density, \emph{n$_e$} (cm$^{-3}$), are obtained: 

\begin{equation}
\Phi=1.24\times10^{10}S_{\nu}T_e^{0.35}\nu^{0.1}\theta^{-2}
\end{equation} 

\begin{equation}
n_{e}=122.21\times\sqrt{\frac{S_{\nu}T_e^{0.35}\nu^{0.1}}{\eta R\theta^2}}
\end{equation}

\noindent where \emph{$S_{\nu}$} is the measured flux density in mJy and $\nu$ is the frequency of the 
observations in GHz. An average \HII~region electron temperature of \emph{T$_{e}$}$\sim$10$^{4}$ K has 
been assumed. The upper limit on the impinging flux is 2.4$\times$10$^{10}$ photons \cmtwo~s$^{-1}$. 
Comparing this to the combined theoretical flux expected at the interface from the O5 and O6.5 stars 
(2.9$\times10^{10}$ photons \cmtwo~s$^{-1}$; Paper I) we find agreement to within 20\%. Therefore our 
observations were only sensitive enough to have detected radio emission from the bright rim if it and the 
ionising stars lie in the same plane in the sky, and if there was no significant attenuation of
the photon flux due to absorption from dust or recombination along the cloud/stars line of sight. 
However, as already stated, the morphology of the PDR suggests that there is some displacement between 
the positions of the ionising stars and the cloud rim, with the stars lying in the background and the 
cloud in the foreground. Taking this into account it would seem that the non-detection of emission 
coincident with the rim is not significant and does not rule out the possibility of a IBL being 
present at the HII region/cloud interface. More sensitive radio observations are required to 
determine whether or not an IBL surrounds the bright rim, as suggested by the presence of the PDR.

The upper limit for the
electron density was calculated to be 378 cm$^{-3}$, considerably above the critical electron density
(\emph{n$_{critical}$}=25 cm$^{-3}$, \citealt{lefloch1997}) necessary for an IBL to develop around 
the cloud rim. Using this value and again assuming a temperature of 10$^4$ K we have estimated an 
upper limit to the external pressure of the cloud, \emph{P$_{ext}$}, using Eqn 3 Paper I,

\begin{equation}
\frac{P_{ext}}{k_B}=2n_eT_e
\end{equation}

\noindent where \emph{k}$_{\rm{B}}$ is Boltzmann's constant. This gives an upper limit to the external 
pressure of $\sim8\times10^6~P/k_B$. The upper limit for the ionised gas pressure compares well to the 
average value of $\sim6\times10^6~P/k_B$  calculate from the measured ionised gas pressures of seventeen 
BRCs reported in Paper I.

\subsection{Star formation within the SFO 79 molecular cloud}

\subsubsection{Morphology of the bright rim and embedded sources} 

There is a remarkable correlation between the morphology of the PDR (blue) and the optical gas (red) as 
seen in Figure~\ref{fig:msx_3colour}. The extended MSX band A emission would suggest that the stars and 
the bright rim do not lie in the same plane of the sky, but that the stars are slightly in the background 
and the cloud in the foreground with the majority of the stars' ionising flux impinging onto the far side 
of the cloud and thus the hidden face of the cloud. This hypothesis is supported by the sharp edge of the 
bright rim and the 8.3 $\mu$m emission that extends through the cloud, presumably due to transmission 
through the cloud from the hidden face. Set back from the bright rim are three thermal sources: a large 
extended source that is coincident with the embedded UC HII region (described previously in Section 3.2), 
a smaller source approximately 2\arcmin~to the northeast; and a similarly sized source approximately 
5\arcmin~directly north of the position of the UC HII region. These have been labelled A, B and C running 
south to north along the edge of the rim (see \emph{upper right panel} Figure~\ref{fig:msx_3colour}).

The largest of the thermal sources A is coincident with both the IRAS point source and the 
UC HII detected by the radio observations. This source is also identified in the MSX point source
catalogue (MSX5C G336.4922-01.4746).  We calculated the source size to be 34\arcsec~(FWHM) from 
emission seen in the band E image positioned at $\alpha$(J2000)=16$^h~40^m~1.4^s$,  $\delta$(J2000)
=-48$^\circ$~51\arcmin~44\arcsec, only slightly larger than the UC HII region. The majority of the 
band E emission emanates from the central region, however, the remaining emission emanates from a 
much larger region surrounding the central source, suggesting the presence of a thick shell of warm 
diffuse dust roughly 2\arcmin~in diameter. 

The second thermal source B is associated with the MSX point source MSX5C G336.5233-01.4643. This source 
was located within the primary beam of our radio observations described in Section 3.2, however, no radio 
emission was detected toward it (to a 3$\sigma$ rms $\sim$16 mJy). This could be due to the limited 
sensitivity of our observations; that this source does not contain an evolved high-mass star, or that 
this source is at an earlier stage of development than source A. Searches of astronomical databases failed 
to associate source B with any catalogued object and an analysis of 2MASS data failed to identify any 
embedded YSOs. This source was not identified in the IRAS point source catalogue, although there is some 
suggestion of a source in the IRAS HIRES 60 $\mu$m image of this region, but it is not readily separable 
from source A, which dominates the image. The source of the thermal emission in this object is therefore 
uncertain and its identification depends upon future higher resolution near/mid IR and radio observations.

Thermal source C is the smallest of the three thermal sources, located at $\alpha$(J2000)=
$16^h~40^m~0.752^s$, $\delta$(J2000)=-48$^\circ$~47\arcmin~2.41\arcsec~and has been identified in both the 
MSX and IRAS point source catalogues, MSX5C G336.5487-01.4240 and IRAS 16362-4841 respectively. An 
analysis of the IRAS fluxes reveals this source has the colour characteristics 
(e.g., log(F$_{25}$)/log(F$_{12}$)=1.5875 and log(F$_{60}$)/log(F$_{12}$)=2.75, where the latter is an 
upper limit) consistent with this source being an UC HII region \citep{wood1989}. It is unclear why this 
source was not included in the SFO catalogue, given its position within the molecular cloud and its IRAS 
fluxes, since it fulfils all the criteria for being a star forming region. Unfortunately source C was 
located $\sim$5\arcmin~from the pointing centre of our radio observations, which places it on the very 
edge of the 6 cm ATCA primary beam, and thus we have no radio data. We consider this source to be a 
potential UC HII region.

\subsubsection{Star formation associated with SFO 79}

There is strong evidence for ongoing star formation toward the position of the embedded IRAS point
source 16362-4845 (thermal source A) within SFO 79. Paper I reported strong 3 and 6 cm emission coincident 
with the position of the embedded IRAS point source leading the authors to suggest that this cloud 
contained a high mass main sequence star. Further observations reported in this paper have confirmed that 
the radio emission is associated with the IRAS point source. We have also resolved the source of the radio 
emission and found it to be consistent with the presence of a deeply embedded UC HII region. The analysis 
of the radio data has pointed to the presence of a high-mass star of spectral class O9 that is driving 
the UC HII region. We consider that the correlation of the UC HII region with the small IR cluster of 
candidate YSOs identified from the 2MASS data is suggestive of small scale sequential star formation 
induced by the expanding ionisation front and the proceeding photoionisation induced shocks that are 
driven into the surrounding molecular material. The identification of these candidate YSOs need to be 
confirmed by deeper photometric observations in \Jm, \Hm, \ks~(and possibly \emph{L} band, as suggested 
by \citealp{meyer1997}). 

The IRAS point source embedded within SFO 79 is located approximately 1.2 pc west of the bright rim, 
coincident with the northern edge of a small knot of emission indicative of a reflection nebula. This 
identification is supported by the $^{13}$CO image which reveals an area of very low emission to the southeast 
of the IRAS position. If this is indeed a reflection nebula it may indicate the presence of a cavity in 
the area to the southeast of the IRAS position in the direction of the IR cluster, possibly excavated by 
the expanding UC HII region or a bipolar outflow from one of the identified YSOs that make up the IR 
cluster. 

The molecular line data have revealed the presence of a dense molecular condensation coincident with the 
IRAS point source and the UC HII region, offset slightly to the northwest ($\sim$10\arcsec). The molecular 
condensation has a mass of $\sim$51 M$_\odot$, an H$_2$ density of $\sim$3$\times$10$^4$ cm$^{-3}$, and a 
global temperature of 20 K. Although this condensation has similar characteristics to protostellar cores 
observed in similar clouds (e.g. Evans 1999) the condensation does not appear to be centrally-condensed, 
possibly indicating the presence of several protostellar cores that remain unresolved by the 30\arcsec\ beam.  

\subsubsection{Could the star formation have been induced?} 

The morphology of the bright rim and the positions of the three thermal sources, located  at a similar 
projected distance from the rim of the cloud, hints at the possibility of a causal link between the two. 
We suggest that the thermal sources are star-forming regions that may have been induced by the propagation 
of a plane parallel shock front into the cloud. A similar morphology has been suggested for the Orion 
bright bar by \citet{omodaka1986} from high resolution CS (\emph{J}=1--0) observations. This idea is 
supported by the association of two UC HII regions (one confirmed and another identified as a potential 
UC HII region) with thermal sources A and C. Given the relatively short lifetime of UC HII regions 
($\sim$$10^5$ yrs) it seems unlikely to find two within the same cloud at the same time, unless there was 
a common casual link. 

To address this question we need to calculate the time the cloud has been illuminated. If we assume 
that the distance between the ionising stars and the  bright rim of the cloud is close to the 
projected distance (4.2 pc) and that the ionisation front expands at the sound speed of 11.4 km 
s$^{-1}$ it would take approximately 3.6$\times10^5$ years for the ionisation front to reach the 
cloud. In estimating this expansion time scale we have assumed a constant expansion speed (11.4 km s$^{-1}$) 
and a density of 10$^2$ \cmthree~for the intervening material between the ionising stars and the rim of 
the cloud. Given that OB stars have a typical lifetime of a few million years, it is possible 
that the cloud has been exposed to the ionising photon flux for a similar length of time. If we assume
a typical shock velocity of 1.4 km s$^{-1}$ for neutral gas with a H$_2$ density of 4$\times10^4$ 
\cmthree~(Paper I, Thompson et al. 2004b), then the shock crossing time to the UC HII region would be 
$\sim$$10^6$ years. Taken together, the time for the ionisation front to reach the cloud rim and the shock 
crossing time are $\sim$1.4$\times10^6$ years, a considerable fraction of the OB stars' lifetime. It is 
therefore highly likely that the UC HII regions postdate the arrival of the shock front and thus could 
have been induced.

Moreover, given that the dynamical timescale of a UC HII region is generally considered to be $\sim$$10^5$ 
years and that the protostellar phase lasts for $\sim$$10^5$ years the time scales involved are all 
consistent with the hypothesis that the expansion of the UC HII region could itself have gone on to 
trigger star formation in the surrounding region. 

\section{Summary and conclusions} 

We have made measurements of the RRL H92$_\alpha$ and radio continuum using the ATCA, and CO observations
with the Mopra telescope, toward the embedded IRAS point source 16362-4845 associated with SFO 79. We have 
combined these observational data with archival data obtained from both the MSX and 2MASS surveys. Strong 
continuum and RRL emission have been detected toward the embedded IRAS source and have allowed the source 
of the emission to be resolved, revealing the presence of a deeply embedded UC HII region. The molecular 
line observations have revealed the presence of a molecular condensation to the northwest of the UC HII 
region. Analysis of MSX images has revealed the presence of two other thermal sources, B and C, to the 
north of the IRAS point source 16362-4845, which was also positionally coincident with thermal source A. 
Thermal source C was found to have IRAS colours consistent with it also being a possible UC HII region. A 
SIMBAD search of the location of thermal source B failed to associate it with any IRAS point source or 
radio emission, therefore the origin of this emission is uncertain.

We draw the following conclusions from our analysis of the data:

\begin{enumerate}
\item We have resolved the source of the spherical radio emission toward the embedded IRAS point source and 
have determined an average electron density of $\sim$1.4$\times10^4$ \cmthree, an emission measure of 
$\sim$1$\times10^7$~cm$^{-6}$ pc and a deconvolved physical size of $\sim$0.1 pc. As this source fulfils 
all the criteria set out by \citet{wood1989} to identify UC HII regions, we therefore confirm the presence 
of a deeply embedded UC HII region possessing a core-halo morphology. Using the integrated radio fluxes and 
making the simple assumption that the observed emission is due to the presence of a single OB star, we have 
determined an upper limit on the spectral type of the OB star that is powering the UC HII region. We 
estimate the spectral type of the OB star to be O9, however, if the UC HII region is being driven by a 
cluster, rather than a single star, the earliest star present would be one or two spectral classes later, 
i.e. a O9.5 or B0 star. A lower limit on the age of the UC HII region has been calculated to be 
$\sim$1$\times10^4$ yrs, with a upper limit on the age set by the presence of the CTTSs, which have a 
typical age of several million years. The CO data have revealed the presence of a dense molecular 
condensation offset $\sim$30\arcsec~north of the peak position of the radio data and the IRAS point source.

\item  We report the unambiguous detection of a total of 10 candidate YSOs, including 8 previously un-reported, 
within a 90\arcsec~radius of the 
IRAS point source. Five of these YSOs form an IR cluster within the boundary of the UC HII region. Included
in this IR cluster is the only protostar identified and all three of the IR sources that have been 
identified as possible protostellar candidates. Moreover, analysis of the extinction corrected reddening 
values would suggest that the IR cluster contains the most reddened YSO candidates detected in the region. 
All this points to the cluster being relatively young, considering the high number of potential protostars
and the other two extremely reddened sources, we estimate the age to be of order of 10$^5$ yrs. The 
correlation of the IR cluster with the UC HII region, and the similarity in ages, leads us to suggest 
that the formation of these YSOs could have been triggered by the expanding ionisation front of the UC 
HII region. The presence of candidate YSOs toward SFO 79 identify this region as an active star forming 
region. 

\item The MSX band E image has revealed three thermal sources within the SFO 79 bright-rimmed cloud; we 
have labelled these three thermal sources A, B and C (see Figure~\ref{fig:msx_3colour} for details). 
Sources A and C are both associated with IRAS and MSX point sources and with radio emission.
We have identified source A as an UC HII region and have found that source C has IRAS 
colour characteristics consistent with that expected from UC HII regions.  All three sources are
located approximately  at the same distance back from the bright rim and lie along a line that follows 
the morphology of the bright rim. Although the alignment of all three thermal sources along the rim of the 
cloud could be purely circumstantial, it strongly suggests that the star formation within could have been 
induced by the passage of a plane parallel shock driven into the cloud by the ionising stars.

\end{enumerate}

\noindent\emph{Acknowledgements.} 

The authors would like to thank the Director and staff of the Paul Wild Observatory, Narrabri, New South 
Wales, Australia for their hospitality and assistance during the Compact Array and Mopra observing runs. 
JSU would also like to thank Demitri Muna and Mark Price for reading an early draft of this manuscript 
and offering some useful suggestions. Finally we would like JSU and LKM are supported by a PPARC doctoral 
studentships and MAT is supported by a PPARC postdoctoral fellowship grant. This research would not have 
been possible without the SIMBAD astronomical database service operated at CDS, Strasbourg, France and
the  NASA Astrophysics Data System Bibliographic Services. The Digitised Sky Survey was produced at the 
Space Telescope Science Institute under U.S. Government grant NAG W-2166. The images of these surveys 
are based on photographic data obtained using the Oschin Schmidt Telescope on Palomar Mountain and the UK 
Schmidt Telescope. The plates were processed into the present compressed digital form with the permission 
of these institutions. This research makes use of data products from the Two Micron All Sky Survey, which 
is a joint project of the University of Massachusetts and the Infrared Processing and Analysis 
Center/California Institute of Technology, funded by the National Aeronautics and Space Administration and
the National Science Foundation. This research also made use of data products from the Midcourse Space 
Experiment.  Processing of the data was funded by the Ballistic Missile Defense Organization with additional 
support from NASA Office of Space Science.  This research has also made use of the NASA/ IPAC Infrared 
Science Archive,  which is operated by the Jet Propulsion Laboratory, California Institute of Technology, 
under contract with the National Aeronautics and Space Administration.

\end{document}